\documentclass[lettersize,journal]{IEEEtran}
\usepackage{amsmath,amsfonts}
\usepackage{algorithmic}
\usepackage{array}
\usepackage[caption=false,font=normalsize,labelfont=sf,textfont=sf]{subfig}
\usepackage{textcomp}
\usepackage{stfloats}
\usepackage{url}
\usepackage{verbatim}
\usepackage{graphicx}
\usepackage{cite}
\usepackage{cite}
\usepackage{amsmath,amssymb,amsfonts}
\usepackage{algorithmic}
\usepackage{graphicx}
\usepackage{diagbox}
\usepackage{textcomp}
\usepackage{xcolor}
\usepackage[numbers, sort&compress]{natbib}
\usepackage{hyperref}
\def\BibTeX{{\rm B\kern-.05em{\sc i\kern-.025em b}\kern-.08em
		T\kern-.1667em\lower.7ex\hbox{E}\kern-.125emX}}
	
\usepackage[caption=false, font=normalsize, labelfont=sf, textfont=sf]{subfig}

\usepackage[ruled,vlined]{algorithm2e}
\usepackage{multirow}

\makeatletter
\hypersetup{
    colorlinks=true,
    linkcolor=blue,
    filecolor=magenta,
    urlcolor=cyan,
}

\DeclareMathOperator*{\argmin}{arg\,min}

\begin{document}

\title{Towards Secrecy-Aware Attacks Against Trust Prediction in Signed Social Networks}

\author{Yulin Zhu, Tomasz Michalak, Xiapu Luo, Xiaoge Zhang, and Kai Zhou\thanks{Yulin Zhu, Xiapu Luo, and Kai Zhou are with the Department of Computing, The Hong Kong Polytechnic University. E-mail: {\tt yulinzhu@polyu.edu.hk, csxluo@comp.polyu.edu.hk, and kaizhou@polyu.edu.hk}. 
\\ 
Xiaoge Zhang is with the Department of Industrial and Systems Engineering, The Hong Kong Polytechnic University. E-mail: {\tt xiaoge.zhang@polyu.edu.hk}. 
\\
Tomasz Michalak is with the University of Warsaw \& Ideas NCBiR, Warszawa, Poland. E-mail: {\tt tpm@mimuw.edu.pl}.}
}



\maketitle

\begin{abstract}
Signed social networks are widely used to model the trust relationships among online users in security-sensitive systems such as cryptocurrency trading platforms, where trust prediction plays a critical role. In this paper, we investigate how attackers could mislead trust prediction by secretly manipulating signed networks. To this end, we first design effective poisoning attacks against representative trust prediction models. The attacks are formulated as hard bi-level optimization problems, for which we propose several efficient approximation solutions. However, the resulting \textit{basic attacks} would severely change the structural semantics (in particular, both local and global balance properties) of a signed network, which makes the attacks prone to be detected by the powerful attack detectors we designed. Given this, we further refine the basic attacks by integrating some \textit{conflicting metrics} as penalty terms into the objective function. The \textit{refined attacks} become secrecy-aware, i.e., they can successfully evade attack detectors with high probability while sacrificing little attack performance. We conduct comprehensive experiments to demonstrate that the basic attacks can severely disrupt trust prediction but could be easily detected, and the refined attacks perform almost equally well while evading detection. 
Overall, our results significantly advance the knowledge in designing more practical attacks, reflecting more realistic threats to current trust prediction models. Moreover, the results also provide valuable insights and guidance for building up robust trust prediction systems.
\end{abstract}



\section{Introduction}
\IEEEPARstart{E}{stablishing} the trust among users is a crucial premise for the security of many online services. 
For example, in prevalent cryptocurrency trading platforms, such as Bitcoin-Alpha \cite{alpha}, it is important for the traders to know and trust each other \textit{before} committing any transactions to avoid various types of scams. Unfortunately, due to the anonymity feature of cryptocurrency systems, traders usually have very limited knowledge about each other. Consequently, it is usually hard to establish trust directly by knowing each other's identity. Instead, a typical approach is to infer or establish trust based on historical data, such as transaction records, user ratings, and so on. Thus, a trader can initiate a transaction only when the predicted trust between her and the other trader is high enough.

In general, we are faced with this problem of \textit{trust prediction} that is crucial for achieving a secure environment for online services. Essentially, trust prediction aims to predict the \textit{unknown} trust relationships between two users based on the \textit{already established} trust among users. This problem of trust prediction has been well investigated in the literature. Specifically, the existing trust is represented as a \textit{signed social network}\footnote{In the rest of this paper, we will interchangeably use ``signed network" and ``signed graph".}, where nodes in the network represent users and edges indicate relationships among users. In particular, each edge is associated with  a positive ($+$) or negative ($-$) sign, indicating the trust or distrust relation between the two corresponding users. Then, given a signed graph, various analytic tools are proposed to predict the trust between users where the edge signs are absent. These tools are based on a wide spectrum of techniques, such as measuring node similarities \cite{derr2018relevance,sse}, traditional machine learning techniques \cite{esp,pole}, and more recently graph neural networks \cite{sdgnn,sgcn}. The common goal of these efforts is to improve the accuracy of trust prediction. 

A fundamental question is then \textit{how reliable and robust the prediction results are, especially in an adversarial environment where attackers can strategically mislead the prediction.} This question is actually well-grounded from several key observations. Firstly, it is known that many analytical tools for various tasks over graphs are vulnerable to attacks, including but not limited to similarity-based link prediction~\cite{zhou2018attacking}, GNN-based node classification~\cite{mettack}, graph-based anomaly detection~\cite{zhu2021binarizedattack}, and Android malware classification~\cite{hrat}. Thus, it is natural to suspect the robustness of trust prediction tools over signed graphs. Secondly, since the prediction result (i.e., trustworthiness) is related to profits, security, reputation, etc., attackers are fully incentivized to mislead the prediction. For instance, a malicious trader would wish to be predicted as trusted by other legitimate traders. Thirdly, attackers indeed have the ability to mislead prediction. This is because the existing trust relationships (i.e., the signed graph) are actually constructed through a data collection process, which could be easily tampered with. For example, a malicious user could compromise some accounts or create fake accounts to give herself high trust ratings. Reflected in the constructed signed graph, some fake positive edges are inserted. That is, attackers can effectively poison the signed graph, which will further cause wrong predictions. Fourthly, very few works have studied the robustness of trust prediction under attack. To the best of our knowledge, our previous work \cite{attacksignsimilarity} initiates the study on attacking trust prediction tools. However, we only focused on manipulating simple similarity metrics, and the main results are about the computational hardness of the attack (plus some simple heuristic attack algorithms). While in this paper, we concentrate more on complex machine-learning-based trust prediction systems and propose effective while secrecy-aware attacking algorithms.

Against this background, we aim to thoroughly investigate to what extent an adversary can mislead trust prediction by formally studying attacks against representative trust prediction tools.
The study of attacks involves several major challenges. Firstly, similar to other graph analytic tasks, trust prediction in signed graphs features a transductive learning setting, where the training and test data reside in a single graph. Consequently, instead of attacking a fixed prediction model (which is easier), the attacks are simultaneously modifying the training process as well as the test data. Mathematically, attacks are formulated as bi-level optimization problems, which are notoriously hard to solve. Secondly, previous research on attacks mainly imposed a budget constraint on the attacker's capability in the hope that the adversarial manipulated graph would not catch the attention of any defender; in other words, the attack would remain secret. Unfortunately, we show that signed graphs contain much richer \textit{structural semantics}, which makes attacks on signed graphs prone to be detected. Thus, to understand the realistic threats of such attacks, there is an urgent need to design mechanisms to enable secret attacks.

To address the first challenge, we adopt two approximation approaches to 
solve the hard bi-level optimization problems for two representative target trust prediction models.
The two approaches are established upon gradient descent, however, we use different approximation methods to estimate the required gradients which are previously hard to compute. 
Specifically, the first approach are model-agnostic in the sense that it treats the graph adjacency matrix as hyperparameters and compute meta-gradients~\cite{mettack} as the approximation, resulting in three specific attack methods: $\mathsf{FlipAttack}$-meta for attacking \textsf{FeXtra} \cite{esp}, and $\mathsf{FlipAttack}$-unsymR and $\mathsf{FlipAttack}$-symR for attacking \textsf{POLE} \cite{pole}. 
The second approach utilizes the specific properties of the target model,
and transform the complex bi-level optimization problem to single-level problem, resulting in an attack method $\mathsf{FlipAttack}$-OLS for attacking \textsf{FeXtra}. These approaches result in the \textbf{basic attacks} against trust prediction where only a budget constraint is considered.

For the second challenge, we firstly develop three attack detectors based on different types of techniques that can distinguish the attacked graphs from clean ones, and secondly propose techniques that can allow attacks to bypass the prior attack detectors, achieving the secrecy of attacks. 
The main idea is to add some meaningful \textit{conflicting metrics} (detailed later) into the attack objective function as penalty terms. As a result, we can enable the new attacks to evade the attack detectors with high probabilities while sacrificing little attack performance. In particular, by adjusting the degree of penalties, we observe a trade-off between the capability of evading detection and attack performance. That is, the \textbf{refined attacks} become \textbf{secrecy-aware}. 

We conduct comprehensive experiments to test the effectiveness of the basic attacks and refined attacks on three real-world signed graphs. Our main contributions are summarized as follows: 
\begin{itemize}
	\item We design several basic attacks against two representative trust prediction models to demonstrate that an adversary could effectively manipulate trust prediction.
	\item By digging into the side effects of basic attacks, we show that those attacks could be detected by our carefully designed detectors, i.e., \underline{M}ulti-\underline{v}iew \underline{S}igned \underline{G}raph \underline{A}nomaly \underline{D}etection (\textsf{MvSGAD}), showing the inefficacy of basic attacks in practice.
	\item By exploring the theories underneath signed graph analysis, we propose techniques to refine basic attacks, showing a trade-off between secrecy and attack performance and reflecting more realistic threats to  trust prediction systems.
\end{itemize}

The rest of the paper is organized as follows. We introduce the related works about trust prediction and adversarial graph analysis in Sec.~\ref{sec-related}. Then, we elaborate on two representative target trust prediction models in Sec.~\ref{sec-target-model}. We formulate the attack problem in Sec.~\ref{sec-problem}. Then, we present the design of basic attacks in Sec.~\ref{sec-basic-attacks} and their corresponding refinements in \ref{sec-secrecy-attacks}. We show the experiment results in Sec.~\ref{sec-experiments} and conclude in Sec.~\ref{sec-conclusion}.

\section{Related Works}
\label{sec-related}
\subsection{Trust Prediction in Signed Graphs}
Compared to the analysis of \textit{unsigned} graphs, the analytic tasks, such as trust prediction over signed graphs, essentially rely on social theories, such as balance theory~\cite{tang2016survey}. Several classes of approaches have been proposed for trust prediction. Among them, the most classical and representative approach, introduced by \citet{esp}, casts trust prediction as a classification problem. Specifically, it identifies some hand-crafted metrics for each node pair and treats those metrics as features, which are fed into a machine-learning model for classification. Since this approach relies on feature extraction, we term it as \textsf{FeXtra} in our paper.
Another class of methods predict the signs based on the similarity between nodes and basically,  more similar node pairs are assigned positive edges and vice versa. These methods differ in the ways of computing node similarities. For example, \citet{derr2018relevance} redesigns some similarity metrics over unsigned graphs to adapt them to signed graphs.  \citet{sse} adopts a spectral clustering algorithm based on the signed Laplacian matrix to construct the embeddings for each node, from which the similarities are computed. \citet{pole} instead utilizes the signed autocovariance similarity matrix which captures both topological and signed similarities for polarized signed graphs. Moreover, several other emerging works~\cite{sdgnn,sgcn} employ the deep-learning framework to learn the latent representations of nodes and treat  trust prediction as a downstream task. Their major techniques involve modifying the learning objectives to incorporate balance theory. Meanwhile, an orthogonal line of works investigates the prediction of trust degrees of \textit{nodes} in a weighted signed graph. For example, \citet{otc} defines two important metrics: goodness and fairness to measure the trustworthiness of individual nodes.

\subsection{Adversarial Graph Analysis} Recently, there is a surge of research efforts on attacking various graph analytic tasks, such as node classification~\cite{nettack,mettack}, link prediction~\cite{zhou2018attacking}, community detection~\cite{li2020adversarial}, graph anomaly detection~\cite{zhu2021binarizedattack}, malware detection~\cite{hrat} and so on. The attack methods can be roughly classified into two categories. The first category of attacks is task-specific: the techniques proposed for solving the optimization problem highly rely on the specific properties of the target model. Representative works include attacks against node similarity~\cite{zhou2018attacking}, and centrality measurements~\cite{waniek2018hiding}. The other category of attacks is based on the gradient-descent method and thus is generally applicable to attack any differentiable machine learning models. Among them, the most representative works~\cite{nettack,mettack} adopt a greedy approach that picks the edge with the largest gradient in each iteration. 
This work is among the first few ones to study attacks in \textit{signed} graphs. In particular, \citet{attacksignsimilarity} studied how to manipulate the signed versions of some simple similarity metrics. As mentioned earlier, their main finding is that attacking these similarity metrics is generally NP-hard. In comparison, we target much more complex machine-learning-based trust prediction systems. Their hardness results actually demonstrate that it is a non-trivial task to attack trust prediction, which also motivates us to study more effective attack algorithms.

\section{Target Models of Trust Prediction}
\label{sec-target-model}
Predicting the mutual trust among users is formally studied as a 
\textit{link classification} problem in the literature. Formally, we represent a signed graph as $\mathcal{G} = (V, E_s, E_o)$, where $V$ denotes the node set, $E_s$ and $E_o$ are the set of edges with and without signs, respectively. In particular, for an edge $e = (u,v) \in E_s$, it is associated with a positive ($+$) or negative ($-$) sign to indicate a trust or distrust relationship between the two nodes $u$ and $v$, respectively. In contrast, with respect to a given edge in $E_o$, it denotes that a relationship between two users is observed while the trust or distrust nature of that relationship is unknown to us. Thus, trust prediction (or link classification) aims to classify all the links in $E_o$ as either positive or negative given $\mathcal{G}$. 

\begin{figure}[t]
\centering
\subfloat[Local-level balance\label{a}]{%
       \includegraphics[width=0.5\linewidth]{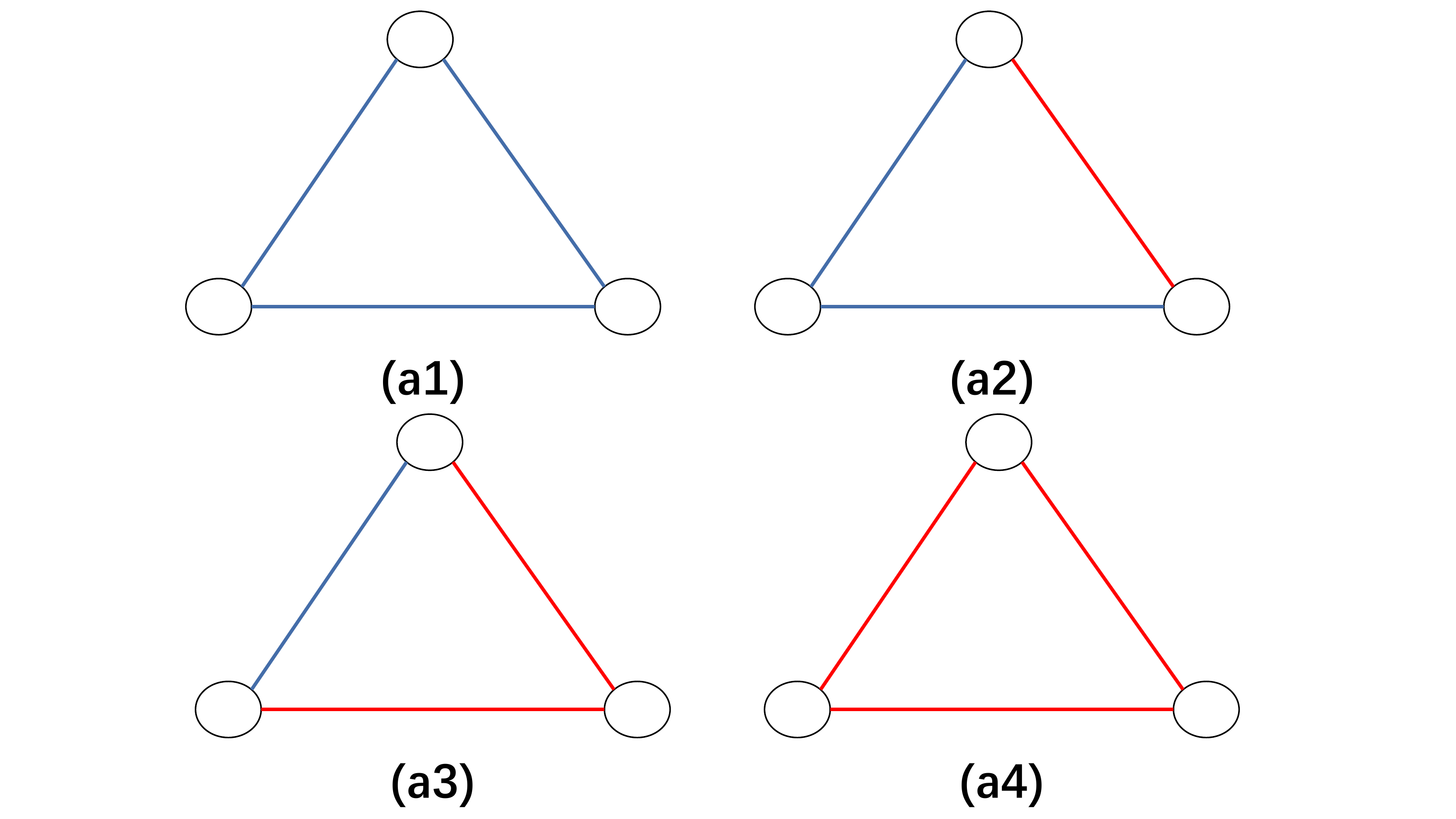}}
\subfloat[Community-level balance\label{b}]{%
       \includegraphics[width=0.5\linewidth]{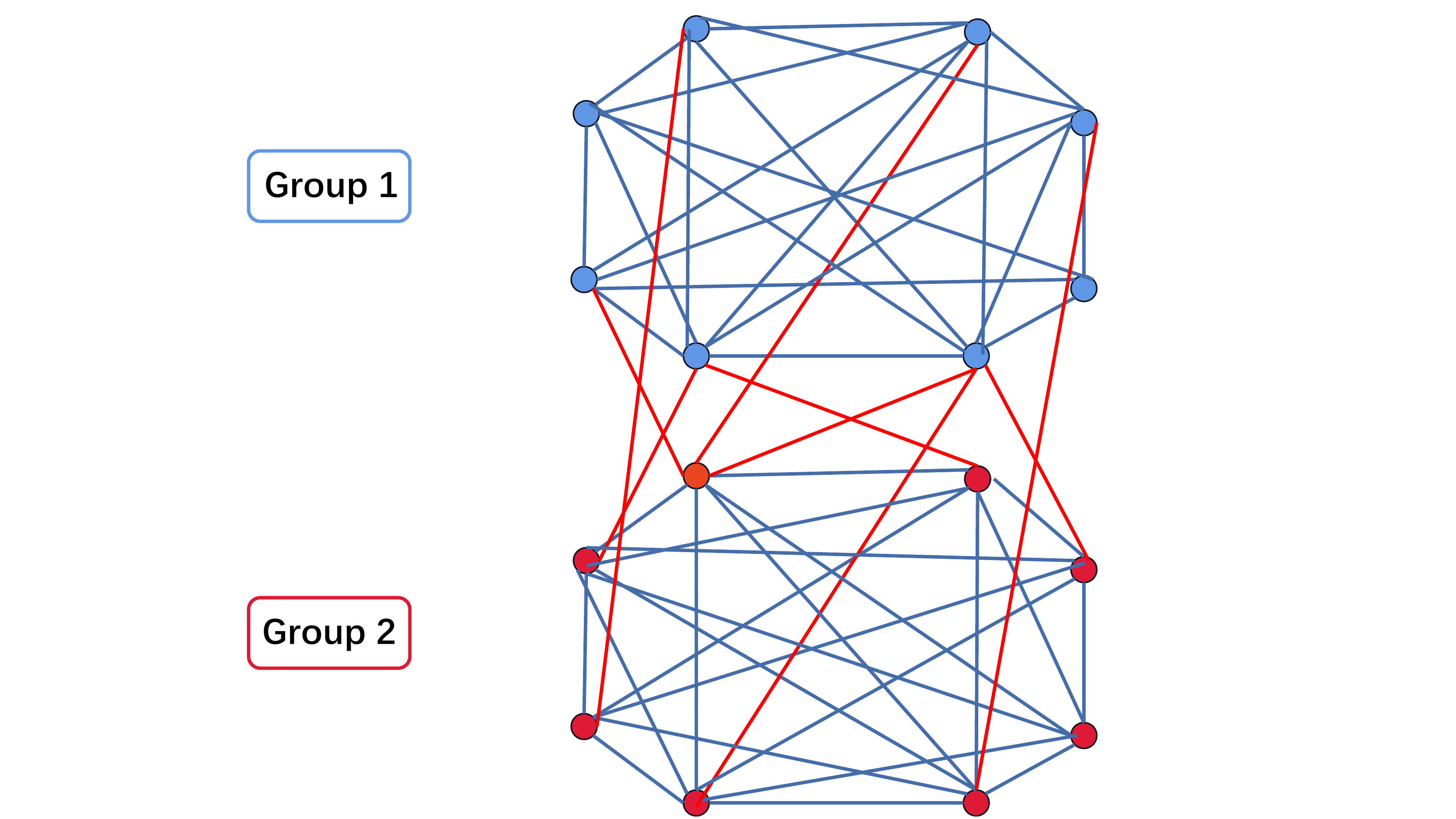}}
\caption{The balance property in signed graphs, where red links represent negative ones. (a): local-level balance. There are four types of closed triads, and a triad is defined as balanced if it contains an even number of negative edges (e.g., (a1) and (a3)). (b): community-level balance. There is a polarization effect at the community level: links within a community tend to be positive while links across communities are likely to be negative.}
\label{fig-structure-balance-polarization}
\end{figure}

While various prediction models have been proposed to tackle this problem, we select two representative ones as the targets of our attacks. The first approach~\cite{esp} (termed \textsf{FeXtra}) is the most classical and widely used one in the literature. In essence, \textsf{FeXtra} extracts hand-crafted features for each edge, and then fed them into a logistic regression model for classification. The second approach~\cite{pole} (termed \textsf{POLE}) employs graph embedding techniques to automatically generate edge embeddings based on which the edges are classified.  These two target models are selected for a few reasons. Firstly, \textsf{FeXtra} is the most classical model that utilizes expert knowledge to identify features with easily interpretable meanings and achieves comparable or even surpass the performance of several other deep learning-based methods (e.g., Table 3 in~\cite{huang2021sdgnn}). \textsf{POLE} serves as a representative of the most recent methods established upon graph representation learning techniques. Secondly, both models integrate balance theory (detailed later in Sec.~\ref{subsec-fex}) into the analysis of signed graphs, however, from different angles. Specifically, \textsf{FeXtra} focuses on the local structure (e.g., closed triads) of a signed graph while \textsf{POLE} examines the structural properties at the community level (see Fig.~\ref{fig-structure-balance-polarization} as an illustrative example). Studying these two models significantly benefits the examination of how attacks would change the balance property from both local and global views. Next, we introduce the necessary details of \textsf{FeXtra} and \textsf{POLE}. 

\subsection{\textsf{FeXtra}}
\label{subsec-fex}
At a high level, for each edge in the graph, \textsf{FeXtra} first extracts some \textit{structural features} and then uses the Logistic Regression (LR) model to compute the probability that a link has a positive sign based on those extracted features. 
The design of those structural features in the LR approach relies on the balance theory \cite{tang2016survey} originated from social science. In fact, resorting to social science theories (e.g., balance theory and status theory) is a unique trait that differentiates the analysis of signed graphs from that of ordinary unsigned ones. Basically, balance theory states that the trust/distrust relationships among a group of three people should be balanced, coinciding with the intuition that ``the enemy of my enemy is my friend". Reflected on the graph structure, a triad is balanced if the number of negative signs over the three edges is even. The hypothesis that all triads in a signed graph should be balanced constitutes the foundation for predicting the edge signs.

Guided by the balance theory, \cite{esp} considered \textit{degree features} and \textit{triad features} associated with each link $(u,v)$.  The degree features are $d^{+}_{u}$, $d^{-}_{u}$, $d^{+}_{v}$ and $d^{-}_{v}$, where $d^+$ and $d^-$ are the numbers of neighbors connected by positive and negative links, respectively. The triad features consider the common neighbors of $u$ and $v$. Specifically, let $\Gamma_{uv}$ be the set of common neighbors. For any triad $\{u,w,v\}$ with a common neighbor $w\in\Gamma_{uv}$, there are four combinations of the signs on the two edges $(u,w)$ and $(w,v)$. Use $\Delta^{++}_{uv}$ to denote the number of triads where both $(u,w)$ and $(w,v)$ have positive signs. Similarly, one can define $\Delta^{+-}_{uv}$, $\Delta^{-+}_{uv}$ and $\Delta^{--}_{uv}$. Thus, any node pair $(u,v)$ can be represented by a nine-dimensional feature vector 
\begin{equation}
	\label{eqn-FExtra-features}
	\mathbf{x}_{uv} = (d_{u}^{+}, d_{u}^{-}, d_{v}^{+}, d_{v}^{-},|\Gamma_{uv}|,\Delta^{++}_{uv},\Delta^{+-}_{uv},\Delta^{-+}_{uv},\Delta^{--}_{uv}).
\end{equation}
Then all the features can be summarized as a matrix $\mathbf{X}^{m\times 9}$, where $m$ denotes the number of links in the signed graph.

Now, predicting signs is a typical supervised classification problem. \citet{esp} employs logistic regression for trust prediction. Specifically, for a link $(u,v)$ with feature vector $\mathbf{x}_{uv}$, the probability that this link has a positive sign is given by:
\begin{equation}
	\label{eqn-logist-regression-prob}
	P((u,v) =+1|\mathbf{x}_{uv})=\frac{1}{1+e^{-(\mathbf{\mathbf{x}_{uv}\theta})}}, 
\end{equation}
where $\theta$ denotes the parameters of the learned LR model. Finally, \textsf{FeXtra} determines $(u,v)$ as positive if $P((u,v) =+1|\mathbf{x}_{uv})>0.5$. For brevity, we denote $f_{\theta}(\cdot)$ as the logistic function with parameter $\theta$.   

\subsection{\textsf{POLE}}
In comparison, \textsf{POLE} looks at the balance property of a signed graph from a global view and investigates an intriguing effect called \textit{polarization}.
Specifically, polarization suggests that a signed graph can be partitioned into two conflicting groups/communities, where nodes within each group are densely connected by positive links while nodes across two groups are connected by negative links. This phenomenon of polarization is most exemplified in politics. For example, the politicians in the U.S. congress naturally form two parties with different political views.

\textsf{POLE}~\cite{pole} utilizes a graph-embedding-based approach, where the links embeddings are generated by a modified random walk process over signed graphs to jointly capture the topological and semantic similarities. 
Specifically, \citet{pole} re-design the random walk process
by adding link signs into the random-walk transition matrix which accumulates the probabilities of a walk from the source node to the target node. 
This results in a better characterization of balance property (in particular, polarization) in signed graphs.
Mathematically, given the adjacency matrix $\mathbf{A}\in\mathbb{R}^{n\times n}$ with entries in $\{+1,-1,0\}$ of a signed graph $\mathcal{G}$, the signed random-walk transition matrix
is calculated as: 
\begin{equation}
	\label{eqn-signed-rw-transition}
	\mathbf{M}(t)=\exp(-(\mathbf{I}-\mathbf{D}^{-1}\mathbf{A})t),
\end{equation}
where $\mathbf{D}=\mathsf{diag}\{\sum_{j=1}^{n}\mathbf{A}_{ij}\}_{i=1}^{n}$ is the degree matrix, $\mathbf{I}\in\mathbb{R}^{n\times n}$ is the identity matrix, and $t$ denotes the length of a walk. 
Next, \textsf{POLE} introduces the signed autocovariance similarity by incorporating node degree information into the signed random-walk to create better links embeddings for trust prediction.
Specifically, the signed autocovariance similarity matrix $\mathbf{R}(t)$ is computed from $\mathbf{M}(t)$ as:
\begin{subequations}
	\label{eqn-signed-autocovariance-similarity}
	\begin{align}
		&\mathbf{R}(t)=\mathbf{M}(t)^{T}\mathbf{WM}(t),\\
		&\text{where} \quad \mathbf{W}=\frac{1}{\sum_{u}d_{u}}\mathbf{D}-\frac{1}{(\sum_{u}d_{u})^{2}}\mathbf{dd}^{T}.
	\end{align}
\end{subequations}
where $\mathbf{W}$ is the weight matrix constructed by node degrees $d_i$. 

Note that the difference of signed and unsigned autocovariance similarity matricies $\mathbf{R}(t)$ lies in that they are computed from the signed adjacency matrix $\mathbf{A}$ and the unsigned version $|\mathbf{A}|$, respectively. To differentiate them, we 
use $\mathbf{R}(t)^{sign}$ and $\mathbf{R}(t)^{abs}$ to represent the signed/unsigned autocovariance similarity matrix, respectively.
For trust prediction, 
To predict trust, 
\textsf{POLE} uses the concatenation of $\mathbf{R}(t)^{sign}$ and $\mathbf{R}(t)^{abs}$ as the 
embedding of a link $(u,v)$. All link embeddings are then treated as features that are fed into a logistic regression model, similar to that of $\textsf{FeXtra}$.

\section{Problem Formulation}
\label{sec-problem}

\subsection{Threat Model}
We consider a scenario that an attacker is capable of manipulating the edge signs in a signed graph, which is subsequently observed by an analyst who will conduct trust prediction over the manipulated graph. Specifically, we denote the original clean graph as $\mathcal{G}^0 = (V^0, E_s^0, E_o^0)$, of which the attacker has a full knowledge. Note that $E_o$ is the set of links whose signs are missing and need to be predicted.  We assume that the ground truth signs for those links remain unknown to \textit{both the attacker and analyst}.

By Kerckhoffs's principle in security, we assume that the attacker has a full knowledge of the prediction model (i.e., white-box attack) with the goal of assessing the robustness of the trust prediction systems in the worst case. In addition, we also test the transferability of our attack in a black-box manner (Sec.~\ref{sec-transfer}).
The attacker's goal is to disrupt the function of trust prediction by maximizing the prediction errors. To this end, the attacker can change (more specifically, flip) the signs of those edges in $E_s^0$. We emphasize that our model and approaches is easily extensible to other attack scenarios, such as erasing or adding the signs. We use $E_s^a$ to denote the  set of edges with changed signs. Consequently, the attacked graph $\mathcal{G}^a = (V^0, E_s^a, E_o^0)$ is observed by the analyst. Finally, based on $\mathcal{G}^a$, the analyst employs various methods for trust prediction.

\subsection{Attack Formulation}
Suppose $\mathbf{X}$ denotes the feature vectors extracted from graph $\mathcal{G}^a$ for all the links (i.e., $E_s^a \cup E_o^0$), $\mathbf{X}$ is then split into two subsets $\mathbf{X}^{tr}$ representing \textit{training links} $E_s^a$ and $\mathbf{X}^{te}$ denoting \textit{testing links} $E_o^0$. Let $\mathbf{y}^{tr}$ and $\mathbf{y}^{te}$ be the corresponding signs of those training and testing links, where $\mathbf{y}^{te}$ is unknown.

The attacker's goal is to maximize the prediction error. In our case, we measure the prediction error as the cross-entropy loss of the predictions for the test links, denoted as $\mathcal{L}_{test} (\mathcal{G}^a, \theta)$, where $\theta$ summaries the parameters of the prediction model. As a result, the attacker aims to maximize this loss function $\mathcal{L}_{test} (\mathcal{G}^a, \theta)$.

We are then faced with an immediate challenge in computing $\mathcal{L}_{test} (\mathcal{G}^a, \theta)$ as it requires the ground truth signs $\mathbf{y}^{te}$ that are unknown to the attacker. To address this issue, we adopt the approach developed by~\citet{mettack}. Basically, as the attacker has access to all the training data, it is possible to predict the signs of the test links prior to the attack. Specifically, the predicted signs $\hat{\mathbf{y}}^{te}$ produced by the prediction method is used in replacement of $\mathbf{y}^{te}$ when computing $\mathcal{L}_{test} (\mathcal{G}^a, \theta)$.

Of particular importance is the fact that the parameter of the prediction model $\theta$ is closely dependent on the graph $\mathcal{G}^a$. More specifically, by manipulating $\mathcal{G}^a$, the attacker actually induces a change to the features $\mathbf{X}^{tr}$ corresponding to the training data. Consequently, the parameter $\theta$ learned from $\mathbf{X}^{tr}$ would also change accordingly with $\mathcal{G}^a$ when the attacker optimizes $\mathcal{G}^a$ -- this emerges as a unique computational challenge in attacking graph-based prediction systems. Mathematically, we formulate the attack as a bi-level optimization problem:
\begin{subequations}
	\label{eqn-bilevel-optim}
	\begin{align}
	\mathcal{G}^{a*}=& \arg\max_{\mathcal{G}^a}\  \mathcal{L}_{test}(\mathcal{G}^a, \theta^*)\label{eqn-bilevel-optim-1}  \nonumber \\
	&\text{s.t.} \quad \theta^{*}=\arg\min_{\theta} \ \mathcal{L}_{train}(\mathcal{G}^a, \theta),\\
	&\qquad ||\mathcal{G}^a - \mathcal{G}^0||\leq B \label{eqn-bilevel-optim-2}.
	\end{align}
\end{subequations}
Note that Eqn. \eqref{eqn-bilevel-optim-1} indicates that the parameter $\theta^*$ is estimated by minimizing a training loss $\mathcal{L}_{train}(\mathcal{G}^a, \theta)$ and Eqn. \eqref{eqn-bilevel-optim-2} imposes a budget constraint on the attacker's ability to be introduced in details later.

\section{Attacks against Trust Prediction}
\label{sec-basic-attacks}
\subsection{Attacking \textsf{FeXtra}}
The remaining task is to solve the bi-level optimization problem, for which we adopt the typical gradient-descent-based method. Yet, we are faced with several challenges. Firstly, to facilitate the computation of gradients, we need to build a differentiable mapping from the loss functions (more specifically, the feature vectors $\mathbf{X}$) to the graph $\mathcal{G}^a$. To this end, we denote the adjacency matrix of the \textit{ground-truth graph} with test signs as $\mathbf{A}$. Note that the entries in $\mathbf{A}$ has three possible values $\{+1,-1,0\}$. After removing the signs of test links, we obtain the clean graph $\mathcal{G}^0$ with adjacency matrix $\mathbf{A}^0$, which is obtained by setting the entries corresponding to test links as $0$. Let $\mathbf{A}^a$ be the adjacency matrix of the attacked graph $\mathcal{G}^a$, which is treated as the variables in our optimization problem and initially, $\mathbf{A}^a = \mathbf{A}^0$.

We split $\mathbf{A}^a$ into a positive matrix $\mathbf{A}^+$ and a negative matrix $\mathbf{A}^-$ to denote the positive and negative signs, respectively. Specifically, $\mathbf{A}^{+}=\sigma(\mathbf{A}^a)$ and $\mathbf{A}^{-}=\mathbf{A}^{+}-\mathbf{A}^a$, where $\sigma(\cdot)$ is the $\mathsf{ReLU}$ function that would set negative entries as $0$. We note that the entries in both $\mathbf{A}^+$ and $\mathbf{A}^-$ now only have two values $\{1,0\}$, where $1$ indicates the existence of a positive or negative link, respectively. Next, we express the features as functions of $\mathbf{A}^a$ (or equivalently $\mathbf{A}^+$ and $\mathbf{A}^-$) as follows:
\begin{subequations}
    \label{eqn-FExtra}
    \begin{align}
&d^{+}_{i}=\sum_{j}\mathbf{A}^{+}[i,j],\  d^{-}_{i}=\sum_{j}\mathbf{A}^{-}[u,v], i = u \textit{ or } v,\\
&|\Gamma_{uv}|=|\mathbf{A}|^{2}[u,v],\\
&\Delta^{++}_{uv}=(\mathbf{A}^{+}\mathbf{A}^{+})[u,v], \ \Delta^{+-}_{uv}=(\mathbf{A}^{+}\mathbf{A}^{-})[u,v],\\
&\Delta^{-+}_{uv}=(\mathbf{A}^{-}\mathbf{A}^{+})[u,v], \  \Delta^{--}_{uv}=(\mathbf{A}^{-}\mathbf{A}^{-})[u,v],
\end{align}
\end{subequations}
where $\mathbf{M}[i,j]$ denotes the entry in the $i$-th row and $j$-th column of the matrix $\mathbf{M}$. Note that the computation of $|\Gamma_{uv}|=|\mathbf{A}|^{2}[u,v]$ relies on the unknown ground-truth graph $\mathbf{A}$. However, we emphasize that only one actually knows the existence of a test link in the graph while only the sign is not known. That is, we can get the absolute value $|\mathbf{A}|$ such that $|\Gamma_{uv}|$ is computable. For the ease of presentation, we summary the mapping as $\mathbf{X} = \mathcal{F}(\mathbf{A}^a)$.

Now, we reformulate the attack problem as:
\begin{subequations}
	\label{eqn-bilevel-optim3}
	\begin{align}
	&\mathbf{A}^{a*}=\arg\max_{\mathbf{A}^{a}} \quad \mathcal{L}_{test}(f_{\theta^{*}}(\mathbf{A}^{a})) \label{eqn-bilevel-obj} \\
	&\text{s.t.} \quad \theta^{*}=\arg\min_{\theta} \quad \mathcal{L}_{train}(f_{\theta}(\mathbf{A}^{a})), \label{eqn-bilevel-optim3-2}\\
	&\quad \quad f_{\theta}(\mathbf{A}^{a})=\frac{1}{1+e^{-(\mathbf{X}^{tr}\theta)}},\\
	&\quad \quad \mathbf{X}^{tr}=\mathcal{F}(\mathbf{A}^{a}), \quad \frac{1}{4}|\mathbf{A}^{a}-\mathbf{A}^{0}|\leq B,
	\end{align}
\end{subequations}

We adopt a \textit{greedy} approach based on gradient descent to solve the above problem. We first relax the integer constraint on $\mathbf{A}^a$ and treat the entries as continuous values. Then, when computing the gradients $\frac{\partial\mathcal{L}_{test}}{\partial\mathbf{A}^a}$ in each iteration, we choose the link with the maximum magnitude of gradient and flip its sign, until a budget $B$ is reached. However, the challenge of this greedy approach lies in computing each gradient $\frac{\partial\mathcal{L}_{test}}{\partial\mathbf{A}^a_{u,v}}$, since obtaining the parameter $\theta^*$ involves a non-differentiable training process. To address this, we introduce two approximating techniques, resulting in two attack methods: $\mathsf{FlipAttack}$-meta and $\mathsf{FlipAttack}$-OLS.

\subsubsection{$\mathsf{FlipAttack}$-meta}

The first method adopts the meta-learning-based attack strategy \cite{mettack} to tackle the difficulty  of computing the gradient of $\mathcal{L}_{test}(f_{\theta}(\mathbf{A}^{a}))$ with respect to $\mathbf{A}^{a}$. Specifically, the method treats $\mathbf{A}^{a}$ as the hyperparameter and compute $\frac{\partial\mathcal{L}_{test}}{\partial\mathbf{A}^{a}}$ by the chain rule, i.e.,
\begin{subequations}
	\label{eqn-metadescent-meta}
	\begin{align}
		&\frac{\partial\mathcal{L}_{test}}{\partial\mathbf{A}^{a}}=\frac{\partial\mathcal{L}_{test}}{\partial f_{\theta^{L}}(\mathbf{X}^{te})}(\frac{\partial f_{\theta^{L}}(\mathbf{X}^{te})}{\partial\mathbf{X}^{te}}\frac{\partial\mathbf{X}^{te}}{\partial\mathbf{A}^{a}}+\frac{\partial f_{\theta^{L}}(\mathbf{X}^{te})}{\partial \theta^{L}}\frac{\partial \theta^{L}}{\partial\mathbf{A}^{a}}),\\
		&\text{where} \quad \frac{\partial \theta^{l+1}}{\partial\mathbf{A}^{a}}=\frac{\partial \theta^{l}}{\partial\mathbf{A}^{a}}-lr\frac{\partial \mathcal{L}_{train}(f_{\theta^{l}}(\mathbf{X}^{te}))}{\partial \theta^{l}\partial\mathbf{X}^{te}}\frac{\partial\mathbf{X}^{te}}{\partial\mathbf{A}^{a}},
	\end{align}
\end{subequations}
$l$ represents the $l$-th iteration in the inner loop. To this end, we firstly use the vanilla gradient descent on the inner loop:
\begin{equation}
	\label{eqn-gradient-descent}
	\theta^{l+1}=\theta^{l}-lr\frac{\partial\mathcal{L}_{train}(f_{\theta}(\mathbf{A}^{a}))}{\partial\theta^{l}}
\end{equation}
for $L$ iterations. We then obtain the meta-gradient $\frac{\partial\mathcal{L}_{test}}{\partial\mathbf{A}^{a}}$ by chaining back to the initial values $\theta^{0}$ following the chain rule in Eqn.~\eqref{eqn-metadescent-meta}. That is, the meta-gradient accumulates the small perturbations of $\theta$ on $\mathbf{A}^{a}$ in the outer loop.  In this way, we can approximately estimate the gradient $\frac{\partial\mathcal{L}_{test}}{\partial\mathbf{A}^{a}}$ and the parameter $L$ controls both the accuracy and computational complexity of estimation.

\subsubsection{$\mathsf{FlipAttack}$-OLS}
The second method relies on replacing the inner optimization problem Eqn.~\eqref{eqn-bilevel-optim3-2} with a closed-form solution. To this end, we approximate the original logistic regression model by linear regression. Then, by OLS estimation \cite{ols}, we directly compute $\theta^*$ as 
\begin{align}
\label{eqn-theta}
\theta^{*}=([\mathbf{1},\ln \mathbf{X}^{tr}]^{T}[\mathbf{1},\ln \mathbf{X}^{tr}])^{-1}[\textbf{1},\ln \mathbf{X}^{tr}]^{T}\ln \mathbf{y}^{tr}. 
\end{align}

By substituting Eqn.~\eqref{eqn-theta} into the objective function \eqref{eqn-bilevel-optim3-2}, we recast the bi-level optimization into a single-level optimization problem, where the gradients $\frac{\partial\mathcal{L}_{test}}{\partial\mathbf{A}^{a}}$ can be directly computed.

Both $\mathsf{FlipAttack}$-meta and $\mathsf{FlipAttack}$-OLS are greedy methods, however, diverging in the approach to compute $\frac{\partial\mathcal{L}_{test}}{\partial\mathbf{A}^{a}}$. In comparison, $\mathsf{FlipAttack}$-meta is a more general approach but is more computationally costly as we have observed. $\mathsf{FlipAttack}$-OLS requires the existence of a close-form solution but is more efficient. The algorithm for $\mathsf{FlipAttack}$-OLS is shown in Alg.~\ref{alg-ols}.

\begin{algorithm}[h]
	\caption{$\mathsf{FlipAttack}$-OLS}
	\label{alg-ols}
	\textbf{Input}: clean signed graph $\mathbf{A}$, budget $B$, self-training signs label $\hat{\mathbf{y}}^{te}$, training link index $tr$ and testing link index $te$, link signs $\mathbf{y}$, \textsf{FeXtra} model $\mathcal{M}$ with parameters $\theta$, link pool $\mathcal{P}=\emptyset$. \\
	\begin{algorithmic}[1] 
		\STATE Let $b=0$, initialize poisoned graph $\mathbf{A}^{a}=\mathbf{A}$; initialize $\theta$ from uniform distribution $\mathcal{U}[0,1]$.
		\WHILE{$b\leq B$}
		\STATE Obtain features $\mathbf{X}=\mathcal{F}(\mathbf{A}^{a})$ from $\mathcal{M}$, then split features as $\mathbf{X}^{tr}=\mathbf{X}[tr]$ and $\mathbf{X}^{te}=\mathbf{X}[te]$.
		\STATE Split signs as $\mathbf{y}^{tr}=\mathbf{y}[tr]$ and $\mathbf{y}^{te}=\mathbf{y}[te]$.
		\STATE Adopt OLS estimation $\theta^{*}$ for $\mathcal{M}$.
		\STATE Compute the attack loss $\mathcal{L}_{test}(f_{\theta^{*}}(\mathbf{A}^{a}))=\sum \hat{\mathbf{y}}^{te}\log(f_{\theta^{*}}(\mathbf{X}^{te}))+(1-\hat{\mathbf{y}}^{te})(1-\log(f_{\theta^{*}}(\mathbf{X}^{te}))).$
		\STATE Compute the gradients $\frac{\partial \mathcal{L}_{test}(f_{\theta^{*}}(\mathbf{A}^{a}))}{\partial \mathbf{A}^{a}}.$
		\STATE Sort the gradients $\frac{\partial \mathcal{L}_{test}(f_{\theta^{*}}(\mathbf{A}^{a}))}{\partial \mathbf{A}^{a}}$ for each link in descending order, the order is $\tau(1)$, $\tau(2)$, ..., $\tau(|tr|)$. 
		\STATE $k=1.$
		\WHILE{the link $e_{\tau(k)}\in\mathcal{P}$}
		\STATE $k\leftarrow k+1.$
		\ENDWHILE
		\STATE Flip the link $e_{\tau(k)}$'s signs to update the poisoned graph $\mathbf{A}^{a}$.
		\STATE $\mathcal{P}\leftarrow \mathcal{P}\cup\{e_{\tau(k)}\}.$
		\ENDWHILE
		\STATE \textbf{return} $\mathbf{A}^{a}$.
	\end{algorithmic}
\end{algorithm}

\subsection{Attacking \textsf{POLE}}
We proceed to the attacks against \textsf{POLE}, where the major challenge is to design a proper attack objective function to capture the adversarial goal of disrupting trust prediction. We note that \textsf{POLE} essentially relies on the \textit{polarized similarity consistency}~\cite{pole}, meaning that node pairs with positive links are more similar than those with negative links. It was shown in~\cite{pole} that the learned signed autocovariance similarity $\mathbf{R}(t)^{sign}$ (i.e., embeddings) could well capture the polarized similarity consistency in that the signs of the entries in $\mathbf{R}(t)^{sign}$ are consistent with the corresponding link signs. Moreover, the magnitude of an entry can be interpreted as the likelihood of the existence of a positive or negative link. Thus intuitively, we can disrupt trust prediction by lowering the \textit{quality} of the learned signed autocovariance similarity $\mathbf{R}(t)^{sign}$.

To this end, we treat an entry in the learned $\mathbf{R}(t)^{sign}$ as the prediction probability of the existences of a positive link, and use cross-entropy loss to measure the prediction error. Then, attacking trust prediction amounts to maximizing the following attack loss:
\begin{equation}
 \label{eqn-symR-loss}
 \mathcal{L}_{test}=\sum_{e=1}^{|E^{test}|}\hat{\mathbf{y}}_{e}\log(\mathbf{P}_{e})+(1-\hat{\mathbf{y}}_{e})\log(1-\mathbf{P}_{e}),
 \end{equation}
where $\hat{\mathbf{y}}$ is the estimated label over test links obtained by the pre-trained trust prediction model, and matrix $\mathbf{P}$ are the prediction probabilities (with each entry $\mathbf{P}_e$ ranging from $0$ to $1$) normalized from $\mathbf{R}(t)^{sign}$, since entries in $\mathbf{R}(t)^{sign}$ have real values. We detail the normalization from $\mathbf{R}(t)^{sign}$ to $\mathbf{P}$ as below.

First, we normalize the entries in $\mathbf{R}(t)^{sign}$ to $[-1,1]$ through the cosine transformation,
resulting in a cosine autocovariance similarity matrix $\mathbf{R}(t)^{sign}_{cos}$. 
The denominator of $\mathbf{R}(t)^{sign}_{cos}$ is computed through matrix factorization of $\mathbf{R}(t)^{sign}$ as follows:
\begin{equation}
	\label{eqn-matrix-factorization}
	\hat{\mathbf{U}}=\argmin\limits_{\mathbf{U}}||\mathbf{UU}^{T}-\mathbf{R}(t)^{sign}||_{2}^{2}.
\end{equation}
Specifically, we use gradient descent to solve \eqref{eqn-matrix-factorization} to obtain the optimal node embeddings $\hat{\mathbf{U}}$. Then, we can reconstruct
the cosine autocovariance similarity $\mathbf{R}(t)^{sign}_{cos}$ as:
\begin{equation}
	\label{eqn-recon-similarity}
	\mathbf{R}(t)^{sign}_{cos}=\mathsf{clamp}(\frac{\mathbf{R}(t)^{sign}}{||\hat{\mathbf{U}}||\cdot||\hat{\mathbf{U}}^{T}||})\in[-1,1],
\end{equation}
where $\mathsf{clamp}(\cdot)$ is a function clipping the input values to $[-1,1]$. Finally, $\mathbf{P}$ is computed as $\mathbf{P}=\frac{\mathbf{R}(t)^{sign}_{cos}+1}{2}$ with entries in $[0,1]$.

Now we can re-write the attack problem as:  
\begin{subequations}
	\label{eqn-pole-optim}
	\begin{align}
		&\mathbf{A}^{a*}=\arg\max_{\mathbf{A}^{a}} \quad \mathcal{L}_{test}(\mathcal{F}(\mathbf{A}^{a}), \hat{\mathbf{U}}) \label{eqn-pole-obj} \\
		&\text{s.t.} \quad \hat{\mathbf{U}}=\arg\min_{\mathbf{U}} \quad ||\mathbf{UU}^{T}-\mathcal{F}(\mathbf{A}^{a})||_{2}^{2}, \label{eqn-pole-optim-2}\\
		&\quad \quad \mathbf{R}(t)^{sign}=\mathcal{F}(\mathbf{A}^{a}), \label{eqn-pole-optim-3}\\
		&\quad \quad \frac{1}{4}|\mathbf{A}^{a}-\mathbf{A}^{0}|\leq B,
	\end{align}
\end{subequations}
where \eqref{eqn-pole-optim-3} describes a differentiable \textsf{POLE} mapping derived from \eqref{eqn-signed-rw-transition} and \eqref{eqn-signed-autocovariance-similarity}.
Now, we are able to use the greedy strategy guided by gradient-descent to solve the above bi-level optimization problem. We use the same idea of $\mathsf{FlipAttack}$-meta to estimate the gradients, i.e.,
	\begin{align}
	\label{eqn-metadescent-symR}
		&\frac{\partial\mathcal{L}_{test}}{\partial\mathbf{A}^{a}}=\frac{\partial\mathcal{L}_{test}}{\partial \mathbf{R}(t)^{sign}_{cos}}(\frac{\partial \mathbf{R}(t)^{sign}_{cos}}{\partial \mathbf{R}(t)^{sign}}\frac{\partial \mathbf{R}(t)^{sign}}{\partial\mathbf{A}^{a}}+\frac{\partial\mathbf{R}(t)^{sign}}{\partial \mathbf{U}^{L}}\frac{\partial \mathbf{U}^{L}}{\partial\mathbf{A}^{a}}),
	\end{align}
where $\frac{\partial \mathbf{U}^{l+1}}{\partial\mathbf{A}^{a}}=\frac{\partial \mathbf{U}^{l}}{\partial\mathbf{A}^{a}}-lr\frac{\partial||\mathbf{U}^{l}(\mathbf{U}^{l})^{T}-\mathbf{R}(t)^{sign}||_{2}^{2}}{\partial \mathbf{U}^{l}\partial\mathbf{A}^{a}}$,
$l$ represents the $l$-th iteration in the inner loop. We term this attack as $\mathsf{FlipAttack}$-unsymR.


We further introduce $\mathsf{FlipAttack}$-symR as an improvement of $\mathsf{FlipAttack}$-unsymR from efficiency perspective. Note that 
the computational bottleneck of $\mathsf{FlipAttack}$-unsymR is the calculation of $\mathbf{M}(t)^{sign}$ in \eqref{eqn-signed-rw-transition}, which involves the time-consuming matrix exponential operation \cite{MatrixExp}. A direct way to speed up this algorithm is to use eigenvalue decomposition \cite{zhang2011matrix} to transform the original matrix exponentiation to the exponential of its eigenvalues, which, however, requires that the target matrix is symmetric. 
Thus, we use a symmetric signed random-walk transition matrix $\mathbf{M}(t)^{sign}_{sym}$ to approximate the original $\mathbf{M}(t)^{sign}$. Specifically, $\mathbf{M}(t)^{sign}_{sym}$ can be computed as follows: 
\begin{subequations}
\label{eqn-sym-srw-transition}
\begin{align}
	\mathbf{M}(t)^{sign}_{sym}&=\exp(-(\mathbf{I}-\mathbf{D}^{-\frac{1}{2}}\mathbf{AD}^{-\frac{1}{2}})t)\\
	&=\mathbf{Q}diag\{e^{\lambda_{1}},e^{\lambda_{2}},...,e^{\lambda_{m}}\}\mathbf{Q}^{T},
\end{align}
\end{subequations}
where $\mathbf{Q}$ is the eigenvector of $-(\mathbf{I}-\mathbf{D}^{-\frac{1}{2}}\mathbf{AD}^{-\frac{1}{2}})t$ and  $\{\lambda_{1},\lambda_{2},...,\lambda_{m}\}$ are the corresponding eigenvalues. This approximation is demonstrated to be beneficial in the experiments: $\mathsf{FlipAttack}$-symR will speed up around $\times4$ in computational time while having comparable attack performance as $\mathsf{FlipAttack}$-unsymR. The algorithm for $\mathsf{FlipAttack}$-symR is shown in Alg.~\ref{alg-symR}.

\begin{algorithm}[h]
	\caption{$\mathsf{FlipAttack}$-symR}
	\label{alg-symR}
	\textbf{Input}: clean signed graph $\mathbf{A}$, budget $B$, inner training iterations $L$, learning rate $lr$, self-training signs label $\hat{\mathbf{y}}^{te}$, training link index $tr$ and testing link index $te$, link signs $\mathbf{y}$, \textsf{POLE} model $\mathcal{M}$ with parameters $\mathbf{U}$, link pool $\mathcal{P}=\emptyset$. \\
	\begin{algorithmic}[1] 
		\STATE Let $b=0$, initialize poisoned graph $\mathbf{A}^{a}=\mathbf{A}$; initialize $\mathbf{U}$ from normal distribution $\mathcal{N}[0,1]$.
		\WHILE{$b\leq B$}
		\STATE Obtain signed autocovariance similarity $\mathbf{R}(t)^{sign}=\mathcal{F}(\mathbf{A}^{a})$ from $\mathcal{M}$ with symmetric signed random-walk transition matrix.
		\STATE $l=1$
		\WHILE{$l\leq$ $L$}
		\STATE $\mathbf{U}^{l+1}\leftarrow \mathbf{U}^{l}-lr\frac{\partial||\mathbf{U}^{l}(\mathbf{U}^{l})^{T}-\mathbf{R}(t)^{sign}||_{2}^{2}}{\partial \mathbf{U}^{l}}.$
		\STATE $l\leftarrow l+1$
		\ENDWHILE
		\STATE Compute the reconstructed cosine autocovariance similarity $\mathbf{R}(t)^{sign}_{cos}=\frac{\mathbf{R}(t)^{sign}}{||\mathbf{U}^{L}||\cdot||(\mathbf{U}^{L})^{T}||},$ and $\mathbf{P}=clamp(\frac{\mathbf{R}(t)^{sign}_{cos}+1}{2}).$
		\STATE Obtain the attack loss $\mathcal{L}_{test}(\mathcal{F}(\mathbf{A}^{a}),\mathbf{U}^{L})=\sum_{e=1}^{|\mathbf{\hat{y}}^{te}|}\hat{\mathbf{y}}^{te}_{e}\log(\mathbf{P}_{e})+(1-\hat{\mathbf{y}}^{te}_{e})\log(1-\mathbf{P}_{e})$.
		\STATE Compute the meta-gradients $\frac{\partial \mathcal{L}_{test}(\mathcal{F}(\mathbf{A}^{a}),\mathbf{U}^{L})}{\partial \mathbf{A}^{a}}.$
		\STATE Sort the meta-gradients $\frac{\partial \mathcal{L}_{test}(\mathcal{F}(\mathbf{A}^{a}),\mathbf{U}^{L})}{\partial \mathbf{A}^{a}}$ for each link in descending order, the order is $\tau(1)$, $\tau(2)$, ..., $\tau(|tr|)$. 
		\STATE $k=1.$
		\WHILE{the link $e_{\tau(k)}\in\mathcal{P}$}
		\STATE $k\leftarrow k+1.$
		\ENDWHILE
		\STATE Flip the link $e_{\tau(k)}$'s signs to update the poisoned graph $\mathbf{A}^{a}$.
		\STATE $\mathcal{P}\leftarrow \mathcal{P}\cup\{e_{\tau(k)}\}.$
		\ENDWHILE
		\STATE \textbf{return} $\mathbf{A}^{a}$.
	\end{algorithmic}
\end{algorithm}

\section{Towards Secrecy-Aware Attacks}
\label{sec-secrecy-attacks}
In this section, we refine the basic attacks towards a secrecy goal. That is, our refined attacks could evade possible detection while remain unnoticeable to a defender, and at the same time preserve satisfactory attack performances.

\subsection{Side Effects of Basic Attacks}
The previous basic attacks manipulate the data to achieve the malicious goal. A natural concern is that the amount of manipulation would be large enough such that the attack would be detected. Most existing studies imposed a budget constraint on attacker's ability and a few considered more complex constraints, such as degree distribution \cite{nettack}, to limit the amount of modification. However, our key observation is that such simple constraints are insufficient to ensure that the modification is unnoticeable to a defender, mainly due to the rich \textit{structural semantics} of graphs (especially, signed graphs). Moreover, $\mathsf{FlipAttack}$ does not change the degree distribution of the signed graphs.

The major theory underneath the analysis of signed graphs is the balance theory, from which a well-accepted hypothesis is that a naturally observed signed graph should be almost balanced. As a result, attacks against signed graph analysis tools should ensure that the modification would not significantly break the balance property of the signed graph. Otherwise, before conducting the analytic task, anyone can reject an attacked graph.

Thus, we investigate how basic attacks would affect the balance property of a signed graph. To this end, we identify some representative metrics to measure the degree of balance from both local and global perspectives.

\subsubsection{Local Structural Balance}
A common method to measure the degree of balance is to count the number of balanced triads. Specifically, a representative metric, termed $T(\mathcal{G})$, is proposed in~\cite{tang2016survey}, which
computes the fraction of balanced triads in a graph $\mathcal{G}$. 
Mathematically, $T(\mathcal{G})$ is calculated from the adjacency matrix as:
\begin{equation}
	\label{eqn-TG}
	T(\mathbf{A})=\frac{Tr(\mathbf{A}^{3})+Tr(|\mathbf{A}^{3}|)}{2Tr(|\mathbf{A}^{3}|)},
\end{equation}
where $Tr(\cdot)$ denotes the trace of a matrix. While there are many variations of $T(\mathcal{G})$, we use it as the representative metric in experiments.

\subsubsection{Global Structural Balance}
As introduced previously, polarization describes balance property of a signed graph from a global view. Specifically,
\citet{pole} introduced both node-level and graph-level metrics to measure the degree of polarization of a signed graph.  
At the node-level, the degree of polarization is defined as the Pearson correlation coefficient between a node's signed and unsigned random-walk transitions:
\begin{equation}
	Pol(u,t)=corr(\mathbf{M}_{u}^{abs}(t),\mathbf{M}_{u}^{sign}(t)), \label{eqn-polarization}
\end{equation}
where $\mathbf{M}^{abs}(t)$ and $\mathbf{M}^{sign}(t)$ are the unsigned and signed random-walk transition matrices calculated from $|\mathbf{A}|$ and $\mathbf{A}$, respectively. Then, a graph-level polarization is defined as the mean value of the polarization degree over all the nodes:
\begin{equation}
	Pol(\mathcal{G},t)=\mathop{mean}\limits_{u\in\mathcal{G}}(Pol(u,t)). \label{eqn-graph-polarization}
\end{equation}

\begin{figure}[t]
	\centering
 \subfloat[$T(\mathcal{G})$\label{1b}]{%
       \includegraphics[width=0.5\linewidth]{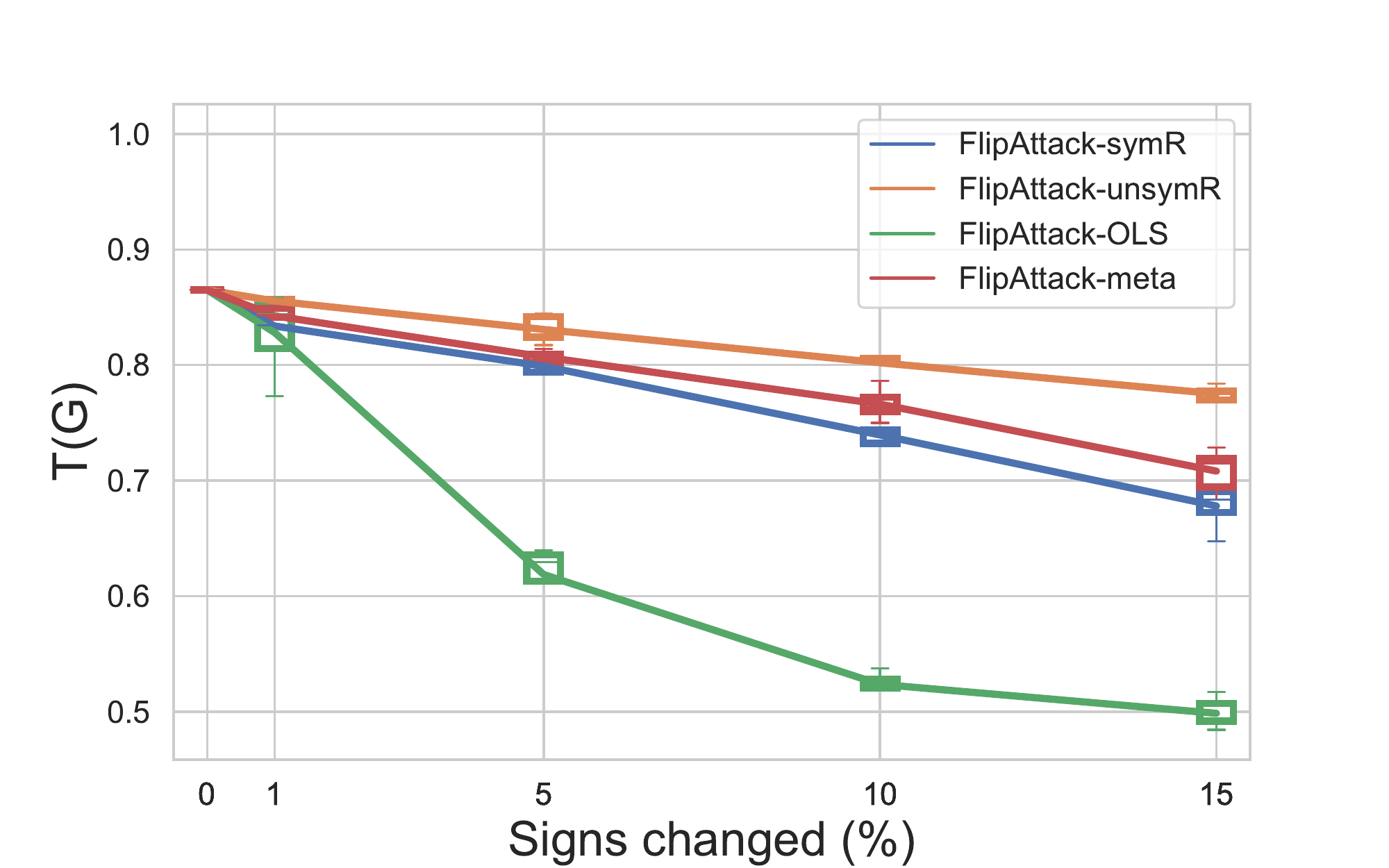}}
	\hfill
  \subfloat[$Pol(\mathcal{G},t)$\label{1b}]{%
       \includegraphics[width=0.5\linewidth]{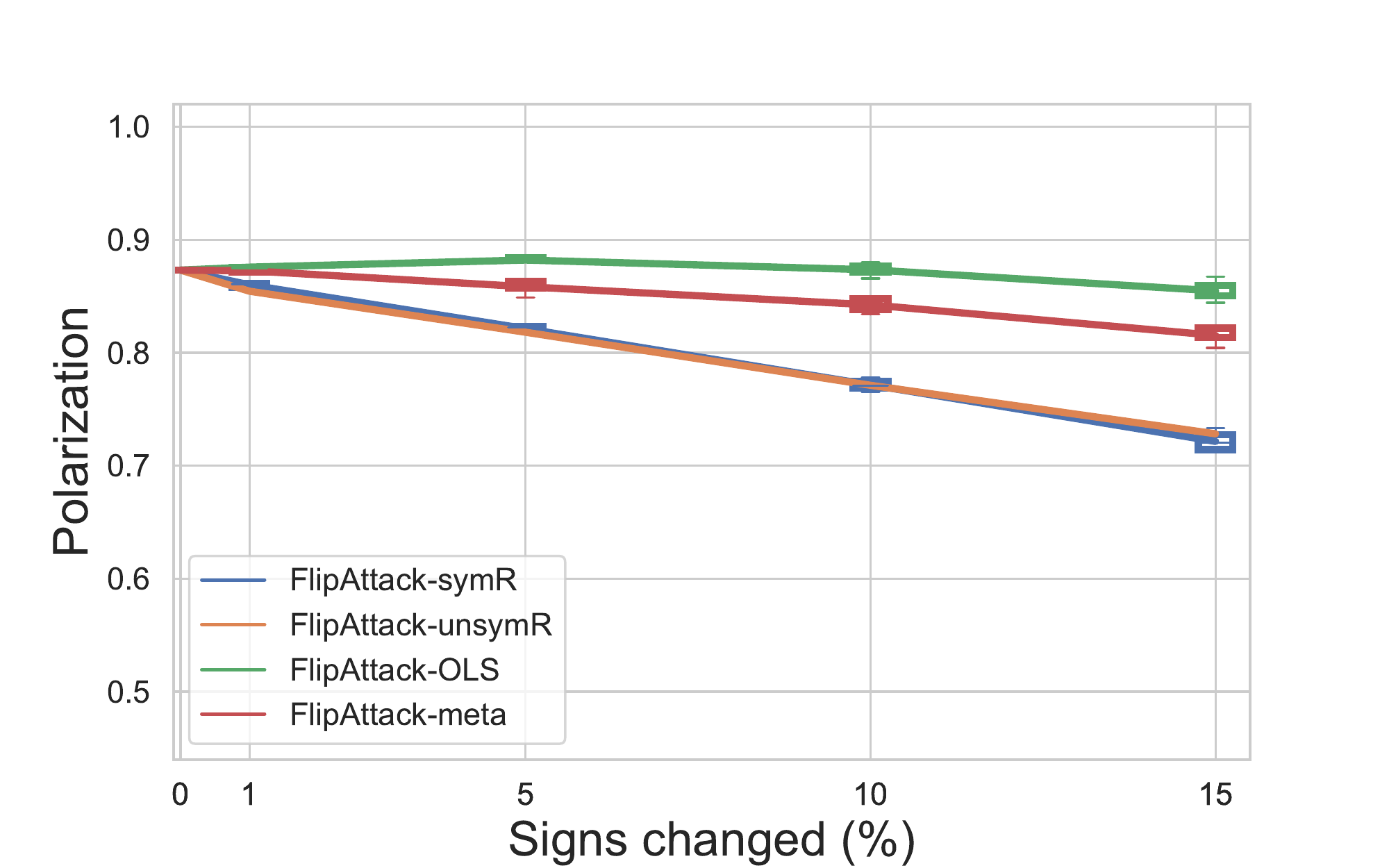}}
	\caption{Changes of $T(\mathcal{G})$ and $Pol(\mathcal{G},t)$ under attacks.}
	\label{fig-balance-under-attack}
\end{figure}

In Fig.~\ref{fig-balance-under-attack}, we show the changes in $T(\mathcal{G})$ and $Pol(\mathcal{G},t)$ under four attacks ($\mathsf{FlipAttack}$-meta, $\mathsf{FlipAttack}$-OLS for attacking \textsf{FeXtra}; $\mathsf{FlipAttack}$-unsymR and $\mathsf{FlipAttack}$-symR for attacking \textsf{POLE}) with increasing attack power on the \textbf{Bitcoin-Alpha} dataset. Notably, even a very small amount of modification ($1\% \sim 5\%$ of the signs) would significantly change the metrics in some cases. This observation is also true on other datasets. That is, \textbf{limiting the amount of modification does not ensure that the attack is unnoticeable}, which provide the motivation for us to consider secrecy-aware attacks. 


\subsection{Attack Detector}
A key step towards achieving secrecy-aware attacks is to anticipate an attack detector employed by a smart defender to detect attacks. We cast this detection problem as an unsupervised graph classification problem (e.g., zero-positive learning \cite{ocsvm}). Specifically, we assume that a defender is able to gather a collection of naturally observed (i.e., clean) signed graphs, possibly from different domains. This assumption reflects the fact that anyone has access to the common knowledge of signed graphs. In our experiments, 
we randomly sample sub-graphs with different node numbers from different datasets to mimic the variety of real-world signed graphs. 

Given this set of clean graphs, we can thus adopt different techniques to train powerful detectors that can differentiate poisoned graphs from clean ones. To ensure the detector is strong and comprehensive enough, we combine three attack detection models with different \textit{views} to capture the anomalous patterns in the signed graphs, 
resulting in an ensemble detector termed \textsf{Multi-view Signed Graph Anomaly Detector} ($\mathsf{MvSGAD}$). In detail, $\mathsf{MvSGAD}$ is composed of three different views: \textsf{Metric-View}, \textsf{TSVD-View} and \textsf{SGCN-View}. 

\subsubsection{\textsf{Metric-View}}
The natural choice is to use the metrics $T(\mathcal{G})$ and $Pol(\mathcal{G},t)$ as features to build a classifier. We use One Class SVM (OCSVM)~\cite{ocsvm} with RBF kernel to implement this idea. 

\subsubsection{\textsf{TSVD-View}}
The second view resorts to graph spectral theory that is tested effective across lots of tasks. We use a spectrum-based embedding method termed \textsf{Truncated Singular Value Decomposition} (\textsf{TSVD}) ~\cite{tsvd} to learn the embedding of the whole graph. It adopts SVD on the signed adjacency matrix: $\mathbf{A}=\mathbf{U\Sigma V}^{T}$, we use $\mathbf{U}\in\mathbb{R}^{N\times d}$ as the node embeddings with dimension $d$. The graph embedding is the mean value of node embeddings in the signed graph. Then, we treat the embedding of a graph as its features, which are fed into OCSVM with RBF kernel to build a detector. 

\subsubsection{\textsf{SGCN-View}}
The third view employs the Signed Graph Convolutional Network (\textsf{SGCN}) \cite{sgcn} as a component to learn the graph embedding, from which an OCSVM is built. Specifically, \textsf{SGCN} integrates balance theory into the message passing process of GCN  to learn node  embeddings. For a node $i$, its embedding is the concatenation of a ``friend" embedding $h_{i}^{B(L)}$ and an``enemy" embeddings $h_{i}^{U(L)}$
where $B(L)$ is the node set $L$-hop away from the center node $i$ along the balance path and  $U(L)$ is the node set along the imbalanced path. Then, we can obtain the graph embedding with MLP augmented with a mean-pooling layer:
\begin{subequations}
	\begin{align}
		&h_{i}^{B(l)}=\text{SGCN}_{W}(h_{i}^{B(l-1)},h_{j}^{B(l-1)},h_{k}^{U(l-1)}|j\in\mathcal{N}_{i}^{+}, k\in\mathcal{N}_{i}^{-}),\\
		&h_{i}^{U(l)}=\text{SGCN}_{W}(h_{i}^{U(l-1)},h_{j}^{U(l-1)},h_{k}^{B(l-1)}|j\in\mathcal{N}_{i}^{+}, k\in\mathcal{N}_{i}^{-}),\\
		&h(G_{k})=\text{MLP}_{W}(\frac{1}{N}\sum_{i=1}^{N}[h_{i}^{B(L)}|h_{i}^{U(L)}]),
	\end{align} 
\end{subequations}
where $\mathcal{N}_{i}^{+}$ and $\mathcal{N}_{i}^{-}$ represent positive and negative neighbors of node $i$, $G_{k}\in\{\mathcal{G}\}_{k=1}^{K}$ contains the clean signed graph and its sub-graphs, MLP$_{W}$ is the fully-connected layer with $\text{ReLU}(\cdot)$ \cite{relu} activation function. Finally, we train the classifier using the following one-class loss \cite{deepsvdd}: 
\begin{align}
	\mathcal{L}_{oc}(\mathcal{G})=\frac{1}{K}\sum_{k=1}^{K}\|h(G_{k})-c\|_{2}^{2}+\frac{\omega}{2}\|W\|_{2}^{2},
\end{align} 
where $W$ contains the parameters in \textsf{SGCN} and the MLP layer. Similar to \cite{deepsvdd}, we fix $c$ to prevent \textit{hypersphere collapse} \cite{deepsvdd}. For convenience, we set $c=[0,0,..,0]^{d}$. After training, the graph embeddings $\{h(G_{k})\}_{k=1}^{K}$ are fed into the kernelized OCSVM to build a detector.

\subsubsection{Ensemble}
In order to comprehensively consider the results from all of the three different views, we can choose the mean, minimum and maximum value of the decision scores of the kernelized OCSVM  in the three classifiers. Intuitively, the mean value means that the ensemble detector uses majority vote to make decisions.
concerns the majority's decision; while 
Choosing the minimum value is a radical strategy, which means if one of the views flags a signed graph as an anomaly, $\mathsf{MvSGAD}$ will treat it as anomalous. In comparison, choosing the maximum value means that $\mathsf{MvSGAD}$ will determine a graph as anomalous only when all views agree, which leads to a conservative strategy. 
We select the normally used AUC score to evaluate $\mathsf{MvSGAD}$'s performance.
, and choose $\mathsf{FlipAttack}$-OLS and $\mathsf{FlipAttack}$-symR on \textbf{Bitcoin-Alpha} dataset as an exemplar. The experimental results are presented in Tab.~\ref{tab-ensemble}. Since $\mathsf{MvSGAD}$ with the max strategy outperforms other strategies for spotting anomalous signed graphs, we choose this strategy to incorporate the three different views in building the ensemble detector.

\begin{table}[h]
	\centering
	\caption{$\mathsf{MvSGAD}$ with different strategies.}
	\label{tab-ensemble}
		\begin{tabular}{|c|c|c|c|}
			\hline
			\diagbox{attack}{AUC}{strategy}&mean&min&max\\
			\hline
			\hline
			$\mathsf{FlipAttack}$-OLS &$0.926$&$0.826$&$\mathbf{0.974}$\\
			$\mathsf{FlipAttack}$-symR&$0.929$&$0.905$&$\mathbf{0.964}$\\
			\hline
		\end{tabular}
\end{table}

\subsection{Refining the Basic Attacks for Secrecy }
We now turn to refining those basic attacks to bypass the previously developed attack detectors.
Intuitively, achieving good attack performance and ensuring unnoticeable attacks are two contradictory goals. More specifically, the former goal will break the balance property of a signed graph and the latter one will preserve the balance property. Our solution is to \textit{quantify this phenomenon}, by identifying some \textit{conflicting metrics}, which attacks and ensuring unnoticeable attacks would change in opposite directions. The metrics to characterize the balance property are a natural choice. 

We thus add the metrics $T(\mathcal{G})$ (for controlling structural balance locally) and $Pol(\mathcal{G},t)$ (for controlling structural balance globally) as penalty terms into the objective function of the optimization problem. That is, we will now simultaneously optimize the original adversarial objective and the penalty terms, corresponding to the joint optimization of the two contradictory goals.  We realize this idea using $\mathsf{FlipAttack}$-OLS and $\mathsf{FlipAttack}$-symR as the examples; however, we emphasize that it can be extended to other attacks
Specifically, for refined attacks, we change the objective function in Eqn.~\eqref{eqn-bilevel-optim3} to 
\begin{equation}
\mathcal{L}_{test}(\mathbf{A}^{a})+\lambda
T(\mathbf{A}^{a}) + \eta Pol(\mathbf{A}^{a},t),
\end{equation}
where we use two hyperparameters $\lambda$ and $\eta$ to adjust the importance of the penalty terms. Intuitively, $\lambda$ and $\eta$ will reflect a trade-off between attack performance and secrecy.

\section{Experiments}
\label{sec-experiments}
In this section, we evaluate our proposed methods through comprehensive experiments from the following key aspects:
\begin{enumerate}
	\item Are basic attacks effective in misleading trust prediction (Section~\ref{sec-basic})?
	\item Can basic attacks be detected (Section~\ref{sec-detect})?
	\item Can refined attacks evade detection (Section~\ref{sec-secret})?
	\item Are attacks still effective against unknown prediction models (i.e., attack transferability, Section~\ref{sec-transfer})?
\end{enumerate}

\subsection{Datasets and Settings}
\begin{figure}[t]
\centering
\subfloat[\textsf{FeXtra} on \textbf{Bitcoin-Alpha}\label{1a}]{%
       \includegraphics[width=0.5\linewidth]{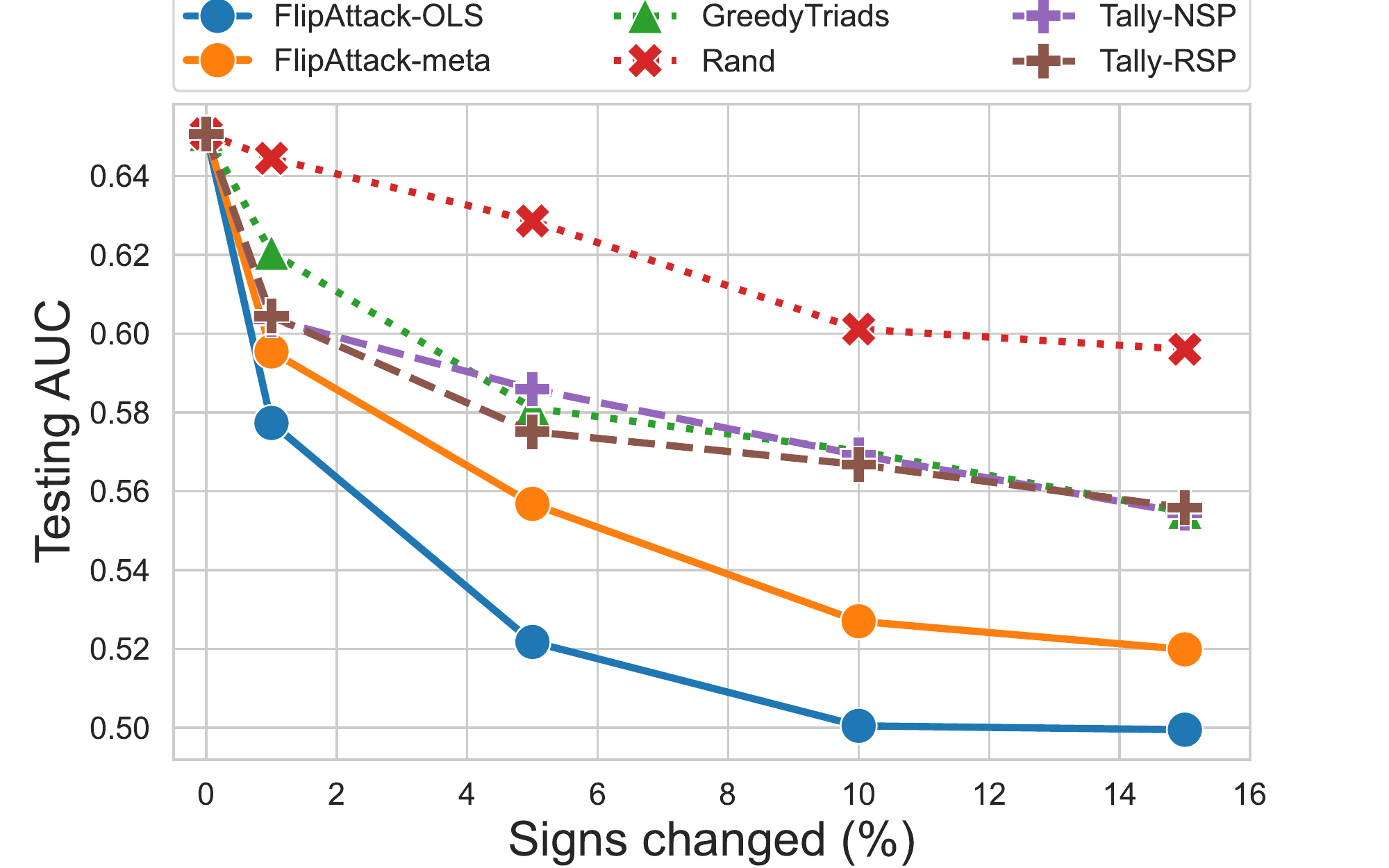}}
 \hfill
 \subfloat[\textsf{FeXtra} on \textbf{Bitcoin-OTC}\label{1b}]{%
       \includegraphics[width=0.5\linewidth]{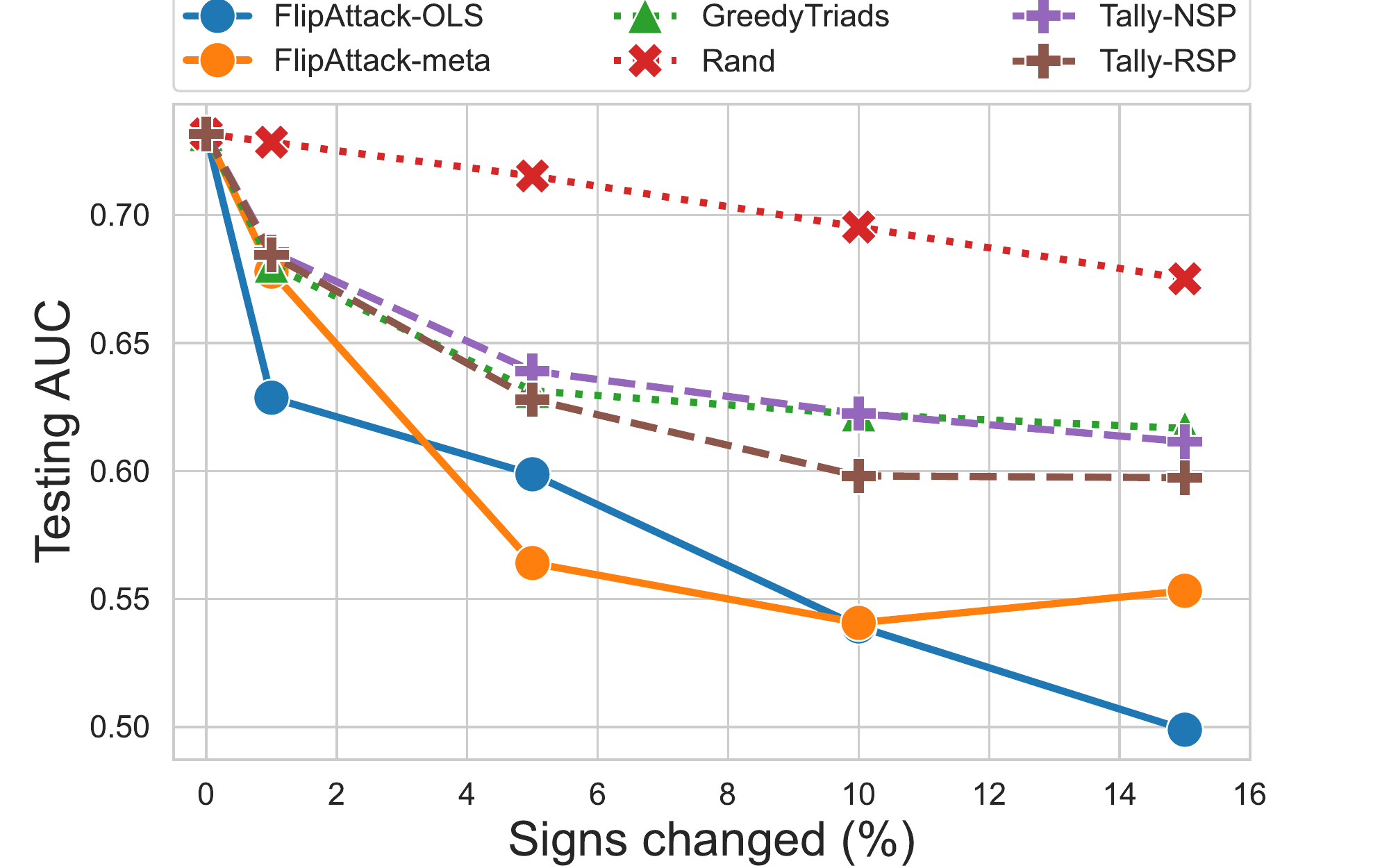}}
	\hfill
\subfloat[\textsf{FeXtra} on \textbf{Word}\label{1b}]{%
       \includegraphics[width=0.5\linewidth]{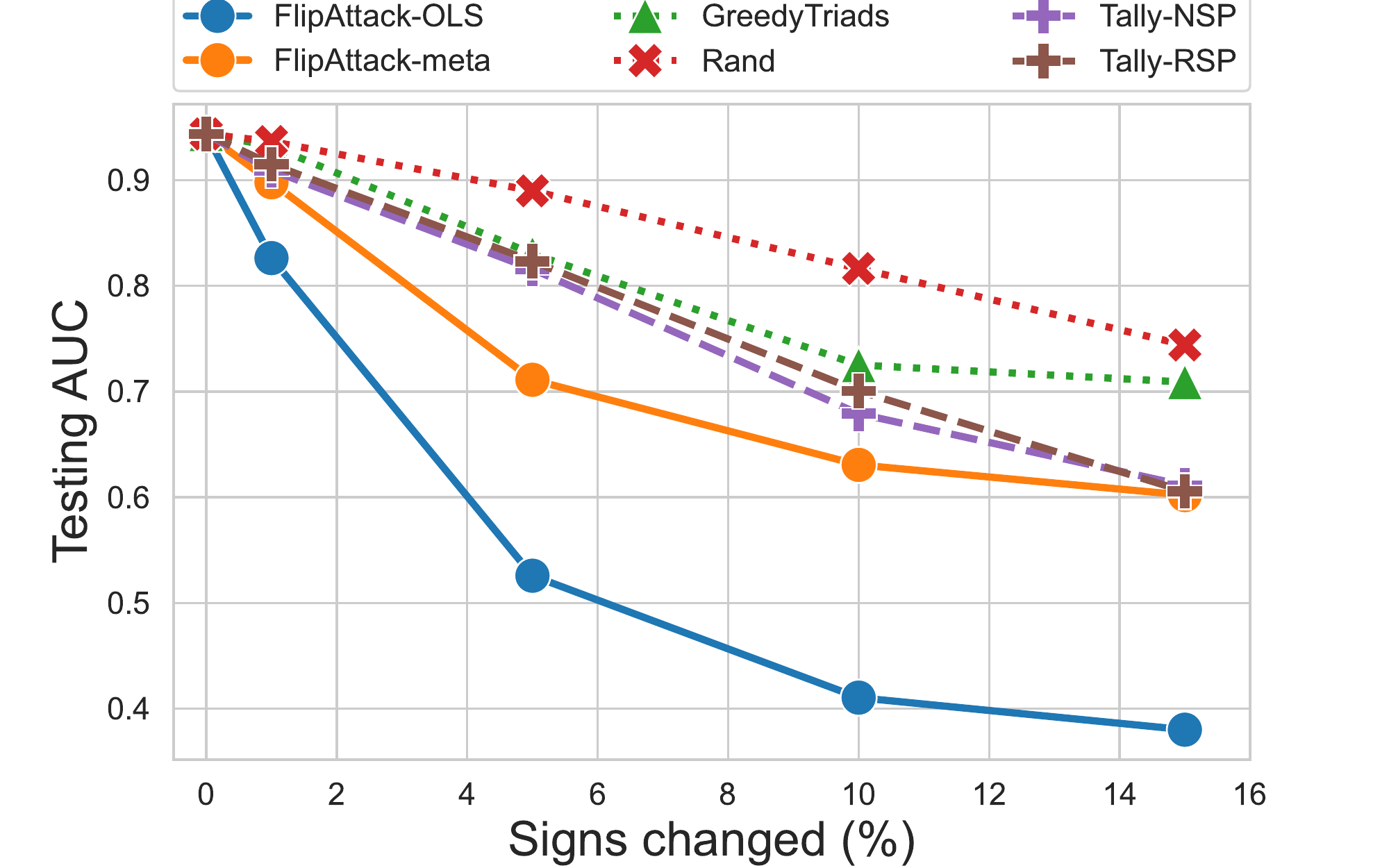}}
	\hfill
 \subfloat[\textsf{POLE} on \textbf{Bitcoin-Alpha}\label{fig-auc-rgraph-sym}]{%
       \includegraphics[width=0.5\linewidth]{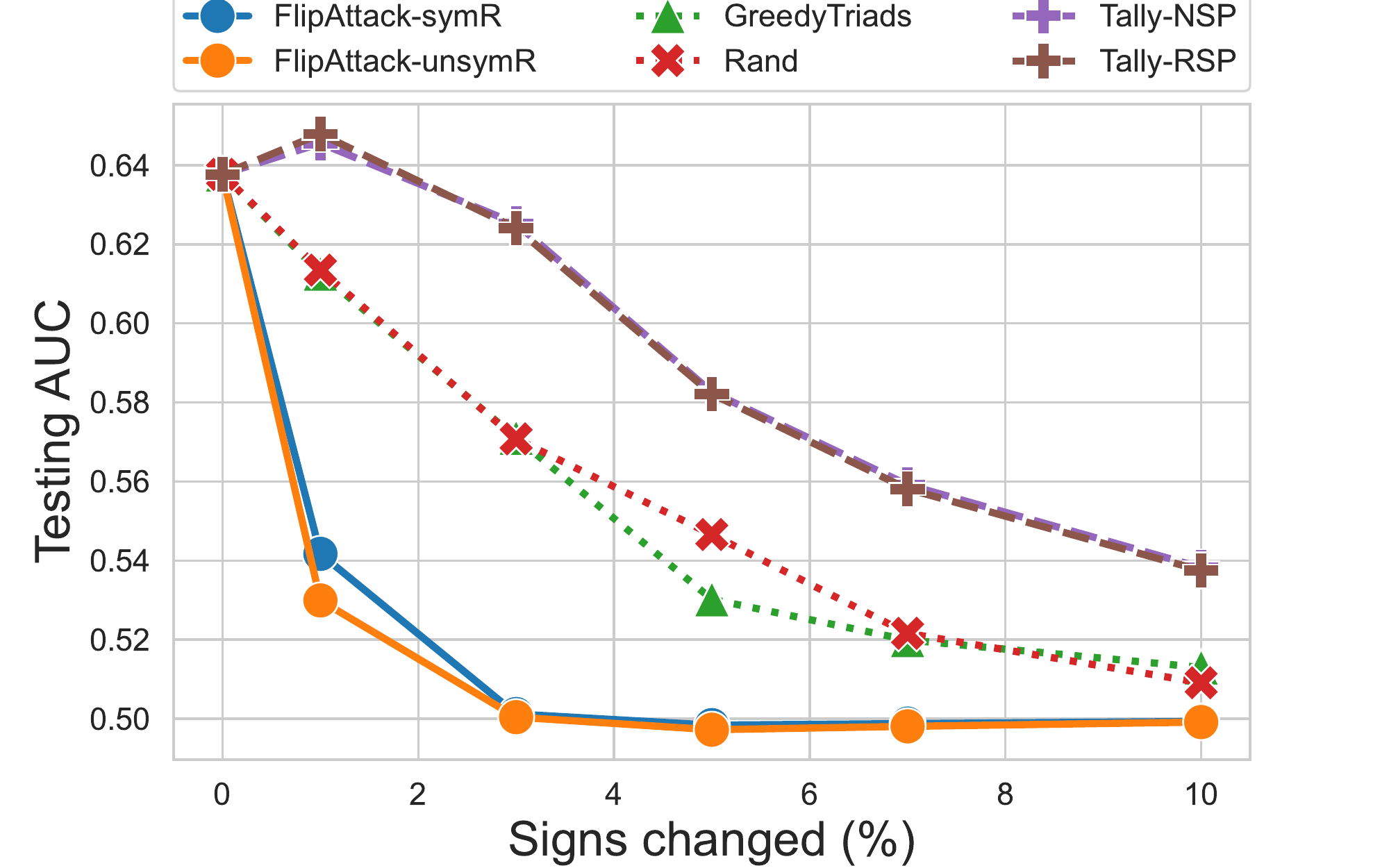}}
	\hfill
  \subfloat[\textsf{POLE} on \textbf{Bitcoin-OTC}\label{2}]{%
       \includegraphics[width=0.5\linewidth]{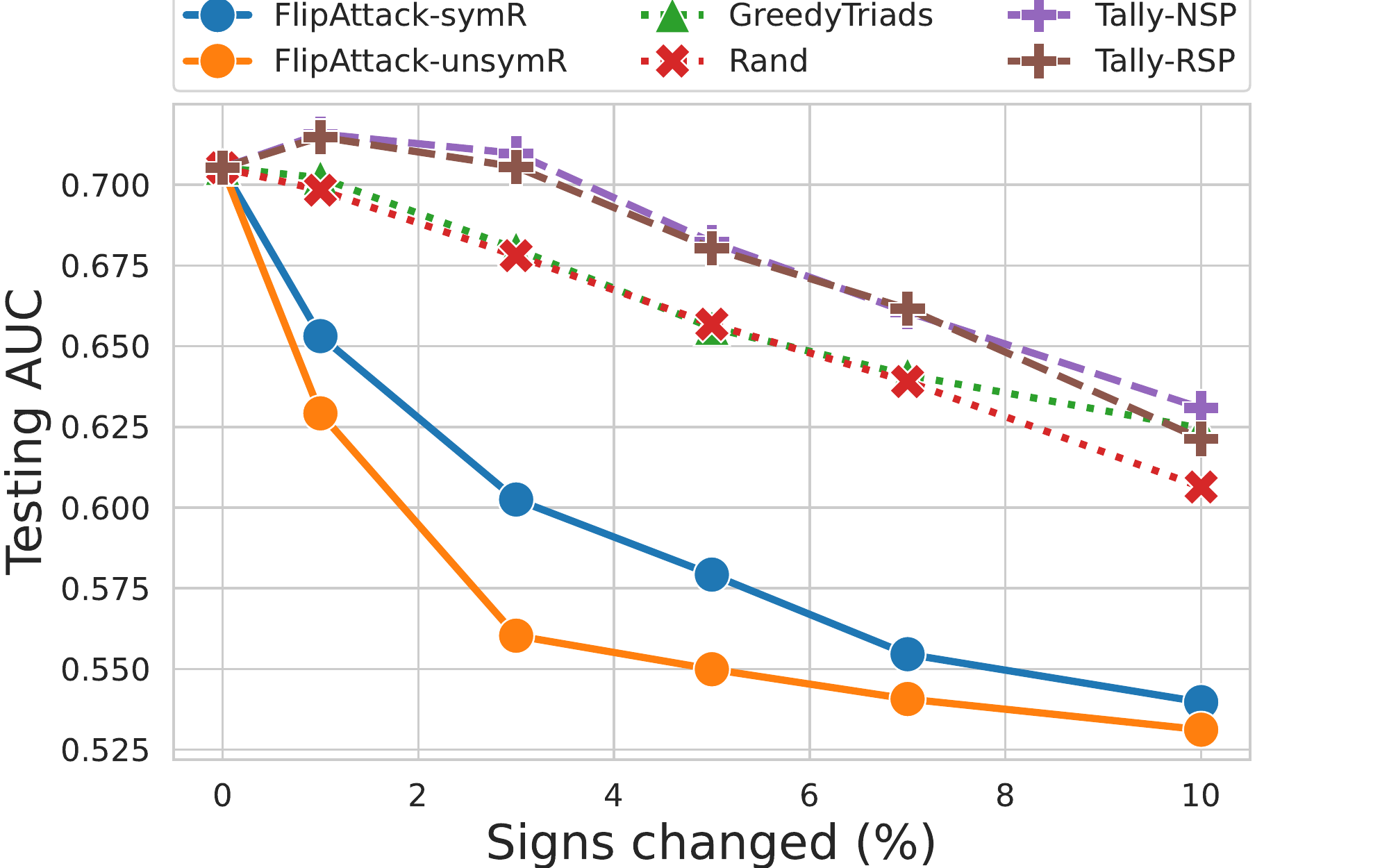}}
	\hfill
\subfloat[\textsf{POLE} on \textbf{Word}\label{3}]{%
       \includegraphics[width=0.5\linewidth]{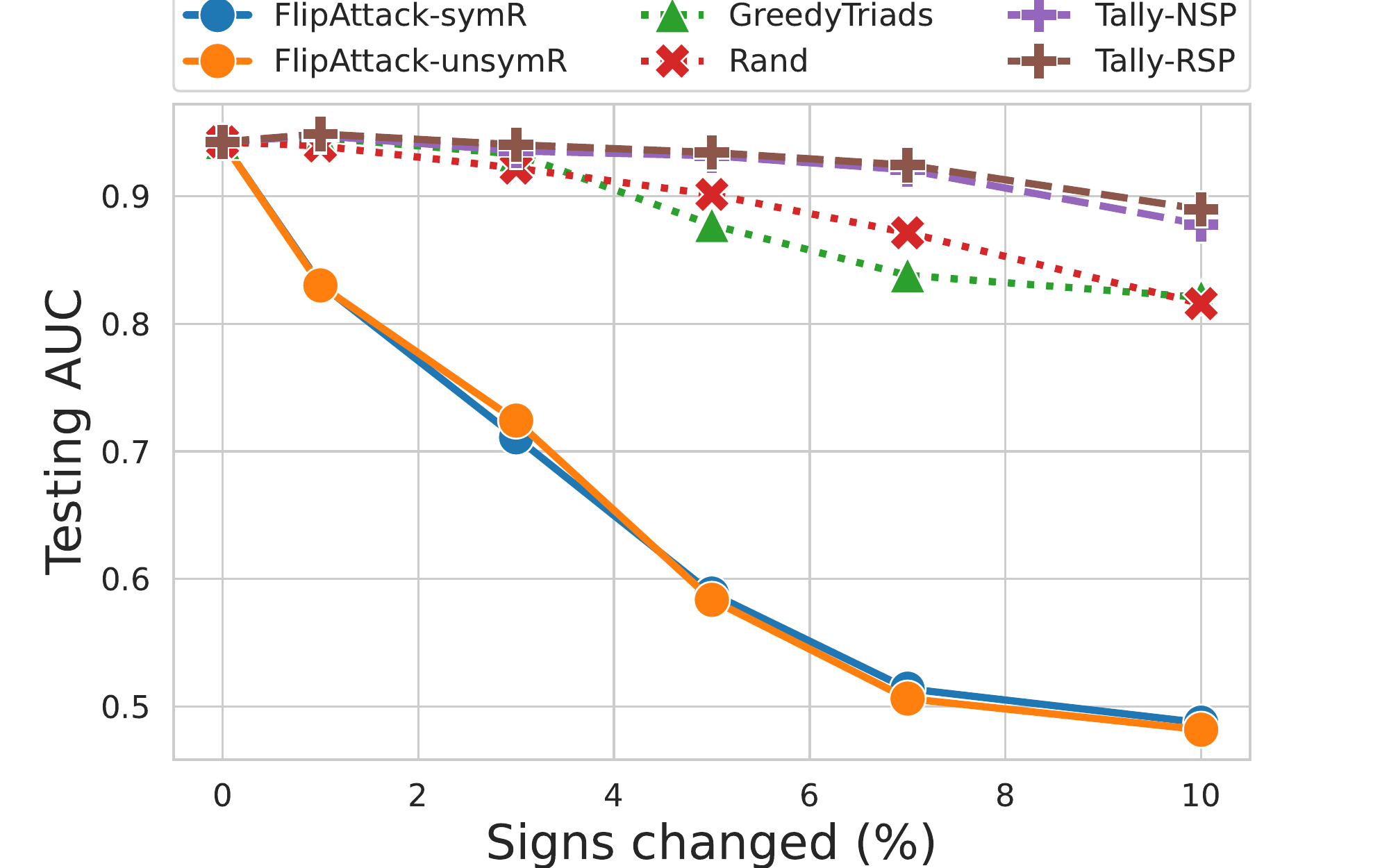}}
	\caption{Testing AUC of \textsf{FeXtra} and \textsf{POLE} under different attacks with various attack powers.}
	\label{fig-auc-rgraph}
\end{figure}

We conduct our experiments on three real-world signed networks: \textbf{Word} \cite{word}, \textbf{Bitcoin-OTC} \cite{otc} and \textbf{Bitcoin-Alpha} \cite{alpha}. We pick out the largest connected component part of the ordinal signed graphs to prevent the singleton structure. For all tasks, the training-test-split is $9:1$. The details of the real-world signed graphs are presented in Tab.~\ref{tab-dataset}. For generating \textbf{Bitcoin-Alpha} sub-graphs, we randomly pick out $1000$, $1500$, $2000$, $2500$, $3000$, $3500$ nodes sub-graphs from \textbf{Bitcoin-Alpha} with 100 times for each scale, we then keep the largest connected component of each sub-graphs. So we totally obtain $600$ sub-graphs for \textbf{Bitcoin-Alpha}. 

\begin{table}[h]
	\centering
	\caption{Statistics of datasets.}
	\label{tab-dataset}
	\begin{tabular}{|c|c|c|c|}
				\hline
				Dataset & $|V|$ & $|E|$ & $|E^{+}|/|E|$\\
				\hline
				\hline
				\textbf{Bitcoin-Alpha} & $3783$ & $24186$ & $0.92$\\
				\textbf{Bitcoin-OTC}   & $5881$ & $35592$ & $0.86$\\
				\textbf{Word}          & $4962$ & $47088$ & $0.80$\\
				\hline
	\end{tabular}
\end{table}

\subsection{Effectiveness of Basic Attacks}
\label{sec-basic}
We compare our attack methods with the following baseline methods:
\begin{itemize}
	\item $\mathsf{Rand}$: It will randomly flip a set of link signs.
	\item $\mathsf{GreedyTriads}$: It will iteratively flip the sign that causes the largest decrease in the number of balanced triangles.
	\item $\mathsf{Tally}$-NSP \cite{attacksignsimilarity}: A heuristic attack method to solve the neutralizing sign prediction problem.
	\item $\mathsf{Tally}$-RSP \cite{attacksignsimilarity}: A heuristic attack method to solve the reversing sign prediction problem.
\end{itemize}

In signed graphs, the number of positive links is much larger than that of negative links, resulting in a very unbalanced dataset. Thus we choose the AUC score to measure the performance of trust prediction. Fig.~\ref{fig-auc-rgraph} shows the average AUC scores on test set under attacks with various attacks powers with 5 independent trials. Specifically, the attack power is measured by the percentage of total signs in the graph. Our key observation is that the four basic attacks $\mathsf{FlipAttack}$-meta, $\mathsf{FlipAttack}$-OLS, $\mathsf{FlipAttack}$-unsymR and $\mathsf{FlipAttack}$-symR are very effective against the two trust prediction models (respectively), and significantly outperform the baseline attacks.
In particular, even with very limited attack power ($< 5\%$), the attacks can severely downgrade the function of trust prediction. 
$\mathsf{FlipAttack}$-OLS outperforms $\mathsf{FlipAttack}$-meta across almost all cases, possibly because $\mathsf{FlipAttack}$-meta uses the general method to estimate the gradients. In comparison, \textsf{POLE} is more sensitive to attacks than \textsf{FeXtra} (note the different scales of the horizontal axis).

\begin{table}[t]
	\centering
	\caption{Time cost (s) and GPU memory usage (MiB) of $\mathsf{FlipAttack}$-unsymR vs $\mathsf{FlipAttack}$-symR and $\mathsf{FlipAttack}$-meta vs $\mathsf{FlipAttack}$-OLS.}
	\label{tab-time-cost}
	\resizebox{0.9\columnwidth}{!}{%
		\begin{tabular}{|c|cc|cc|cc|}
			\hline
			\multirow{2}*{\diagbox{attack}{dataset}}&\multicolumn{2}{c|}{\textbf{Bitcoin-Alpha}}&\multicolumn{2}{c|}{\textbf{Bitcoin-OTC}}&\multicolumn{2}{c|}{\textbf{Word}}\\
			&Time&Mem&Time&Mem&Time&Mem\\
			\hline
			\hline
			$\mathsf{FlipAttack}$-unsymR &$23$&$10781$&$87$&$22711$&$48$&$16815$\\
			$\mathsf{FlipAttack}$-symR   &$6$ &$3469$ &$21$&$6071$ &$14$&$4963$\\
			\hline
			$\mathsf{FlipAttack}$-meta   &$4$   &$4411$ &$7$&$6923$ &$13$&$6141$\\
			$\mathsf{FlipAttack}$-OLS    &$0.5$ &$4467$ &$1$&$6977$ &$1$ &$6195$\\
			\hline
		\end{tabular}
	}
\end{table}

We further collect the time costs and GPU memory usage of $\mathsf{FlipAttack}$-unsymR vs $\mathsf{FlipAttack}$-symR and $\mathsf{FlipAttack}$-meta vs $\mathsf{FlipAttack}$-OLS for conducting one perturbation on different datasets. All the experiments are run on NVIDIA Geforce RTX 3090 GPU. The results are shown in Tab.~\ref{tab-time-cost}. It shows that by using eigenvalue decomposition to replace the matrix exponentiation can significantly speed up the attack method for more than three times as well as decreasing the GPU memory usage around more than three times. On the other hand, although the close form solution slightly occupies more memory usage than meta-learning, it significantly speeds up the attacking algorithm, and the speed gap increase as the graph level increase. We note that since $\mathsf{FlipAttack}$-symR has a comparable performance with $\mathsf{FlipAttack}$-unsymR as shown in Fig.~\ref{fig-auc-rgraph-sym} while $\mathsf{FlipAttack}$-symR is more efficient,  we use it as the representative attack against \textsf{POLE} in later experiments.

\subsection{Detection of Attacks}
\label{sec-detect}
In our experiment, we choose $\mathsf{FlipAttack}$-OLS and $\mathsf{FlipAttack}$-symR as two target attack methods as they will break the balance property ($T(G)$ and $Pol(\mathcal{G},t)$) more severely and needs to be refined. We use the Adam optimizer \cite{adam} with the learning rate equal to $0.001$ and $\omega=10^{-5}$ to train the \textsf{SGCN-View}. The embedding dimension $d$ for \textsf{TSVD-View} and \textsf{SGCN-View}'s ``friend" and ``enemy" embeddings are set as $32$. For each view in $\mathsf{MvSGAD}$, we set the parameter $\gamma$ in the RBF kernel as $\gamma=0.1$. During the training phase, we feed all the $600$ normal signed graphs into $\mathsf{MvSGAD}$ and obtain high-quality graph embeddings with different views for each normal sample and get the corresponding decision score. For testing, we feed all the $25$ poisoned graphs ($5$ poisoned graphs with $5$ different attack powers. For example, the 5 attacking powers for $\mathsf{FlipAttack}$-OLS are $1\%$, $5\%$, $10\%$, $15\%$, $20\%$ while for $\mathsf{FlipAttack}$-symR are $1\%$, $3\%$, $5\%$, $7\%$, $10\%$) into $\mathsf{MvSGAD}$ and obtain decision scores for poisoned graphs. For evaluation, we use AUC scores by comparing the min-max normalization decision scores for normal and poisoned graphs with their true labels ($+1$ for the normal sample and $-1$ for the anomaly sample). Basically, a larger AUC score shows that the detector has a better detection performance.

\begin{table}[h]
	\centering
	\caption{The AUC scores of \textsf{Metric-View}, \textsf{TSVD-View}, \textsf{SGCN-View} and ensemble learning \textsf{MvSGAD} on \textbf{Bitcoin-Alpha}. The AUC score $\tau$ under medium level attacking power $10\%$ is used for $\mathsf{FlipAttack}$-OLS.}
	\label{tab-detect-OLS}
	\resizebox{1.0\columnwidth}{!}{%
		\begin{tabular}{|c|c|c|c|c|c|c|}
			\hline
			$\lambda$ & $\eta$ & $\tau$ & \textsf{Metric-View} & \textsf{TSVD-View} & \textsf{SGCN-View} & \textsf{MvSGAD}\\
			\hline
			\hline
			$0$   &$0$   &$0.50$&$0.982$         &$0.781$         &$0.787$         &$0.974$\\
			$0.01$&$0$   &$0.50$&$0.978$         &$0.753$         &$0.716$         &$0.891$\\
			$0.1$ &$0$   &$0.50$&$0.972$         &$0.786$         &$0.626$         &$0.708$\\
			$1.$  &$0$   &$0.52$&$0.947$         &$0.772$         &$0.642$         &$0.708$\\
			$2.$  &$0$   &$0.53$&$0.946$         &$0.738$         &$\mathbf{0.611}$&$\mathbf{0.667}$\\
			$5.$  &$0$   &$0.53$&$\mathbf{0.787}$&$\mathbf{0.675}$&$0.762$         &$0.801$\\
			\hline
			$0$   &$0.01$&$0.50$&$0.997$         &$0.799$         &$0.828$         &$0.891$\\
			$0$   &$0.1$ &$0.50$&$0.993$         &$0.681$         &$0.719$         &$0.708$\\
			$0$   &$1.$  &$0.50$&$0.997$         &$0.533$         &$0.743$         &$0.708$\\
			$0$   &$2.$  &$0.50$&$0.997$         &$\mathbf{0.532}$&$0.734$         &$\mathbf{0.667}$\\
			\hline
		\end{tabular}
	}
\end{table}

\begin{table}[h]
	\centering
	\caption{The AUC scores of \textsf{Metric-View}, \textsf{TSVD-View}, \textsf{SGCN-View} and ensemble learning \textsf{MvSGAD} on \textbf{Bitcoin-Alpha}. The AUC score $\tau$ under medium level attacking power $5\%$ is used for $\mathsf{FlipAttack}$-symR.}
	\label{tab-detect-symR}
	\resizebox{1.0\columnwidth}{!}{%
		\begin{tabular}{|c|c|c|c|c|c|c|}
			\hline
			$\lambda$ & $\eta$ & $\tau$ & \textsf{Metric-View} & \textsf{TSVD-View} & \textsf{SGCN-View} & \textsf{MvSGAD}\\
			\hline
			\hline
			$0$   &$0$    &$0.50$&$0.940$         &$0.873$         &$0.856$         &$0.964$\\
			$0$   &$0.001$&$0.50$&$0.902$         &$0.866$         &$0.836$         &$0.915$\\
			$0$   &$0.01$ &$0.50$&$0.907$         &$0.862$         &$0.844$         &$0.916$\\
			$0$   &$0.1$  &$0.50$&$0.921$         &$0.861$         &$0.787$         &$0.862$\\
			$0$   &$1.$   &$0.50$&$0.981$         &$0.815$         &$0.610$         &$0.676$\\
			$0$   &$2.$   &$0.50$&$0.996$         &$\mathbf{0.752}$&$\mathbf{0.508}$&$\mathbf{0.590}$\\
			\hline
			$0.01$&$0$    &$0.50$&$0.896$         &$0.872$         &$0.829$         &$0.890$\\
			$0.1$ &$0$    &$0.50$&$0.780$         &$0.875$         &$0.794$         &$0.853$\\
			$1.$  &$0$    &$0.50$&$0.796$         &$0.869$         &$0.780$         &$0.849$\\
			$2.$  &$0$    &$0.50$&$\mathbf{0.460}$&$0.892$         &$0.722$         &$\mathbf{0.545}$\\
			\hline
		\end{tabular}
	}
\end{table}
The first rows (where $\lambda = \eta =0$) in Table \ref{tab-detect-OLS} and Table \ref{tab-detect-symR} shows the performance of different detectors in detecting $\mathsf{FlipAttack}$-OLS and $\mathsf{FlipAttack}$-symR, respectively. One key observation is that the detector \textsf{Metric-View} which relies on metrics of balance is very effective in detecting attacks. It actually demonstrates that the two metrics we identified indeed capture how attacks would change the structural semantics of signed graphs.
Next, we describe how secrecy-aware attacks can evade anomaly detection by tuning different penalties.

\subsection{Towards Secrecy-aware Attacks}
\label{sec-secret}
In our experiment, we adjust the relative importance of preserving the attack performance and evading attack detection via two parameters $\lambda$ and $\eta$. While choosing different $\lambda$ and $\eta$, we want to evaluate the refined attacks from two aspects: how well they can preserve attack performance and how successful they can evade anomaly detection. In particular, we show that $T(\mathcal{G})$  and $Pol(\mathcal{G},t)$ have different effects on evading those three detectors; however, by properly choosing $\lambda$ and $\eta$, the detectors can be successfully evaded while sacrificing little attack performance.

\begin{figure}[!t]
	\centering
    \subfloat[\textsf{FeXtra}\label{1}]{%
       \includegraphics[width=0.5\linewidth]{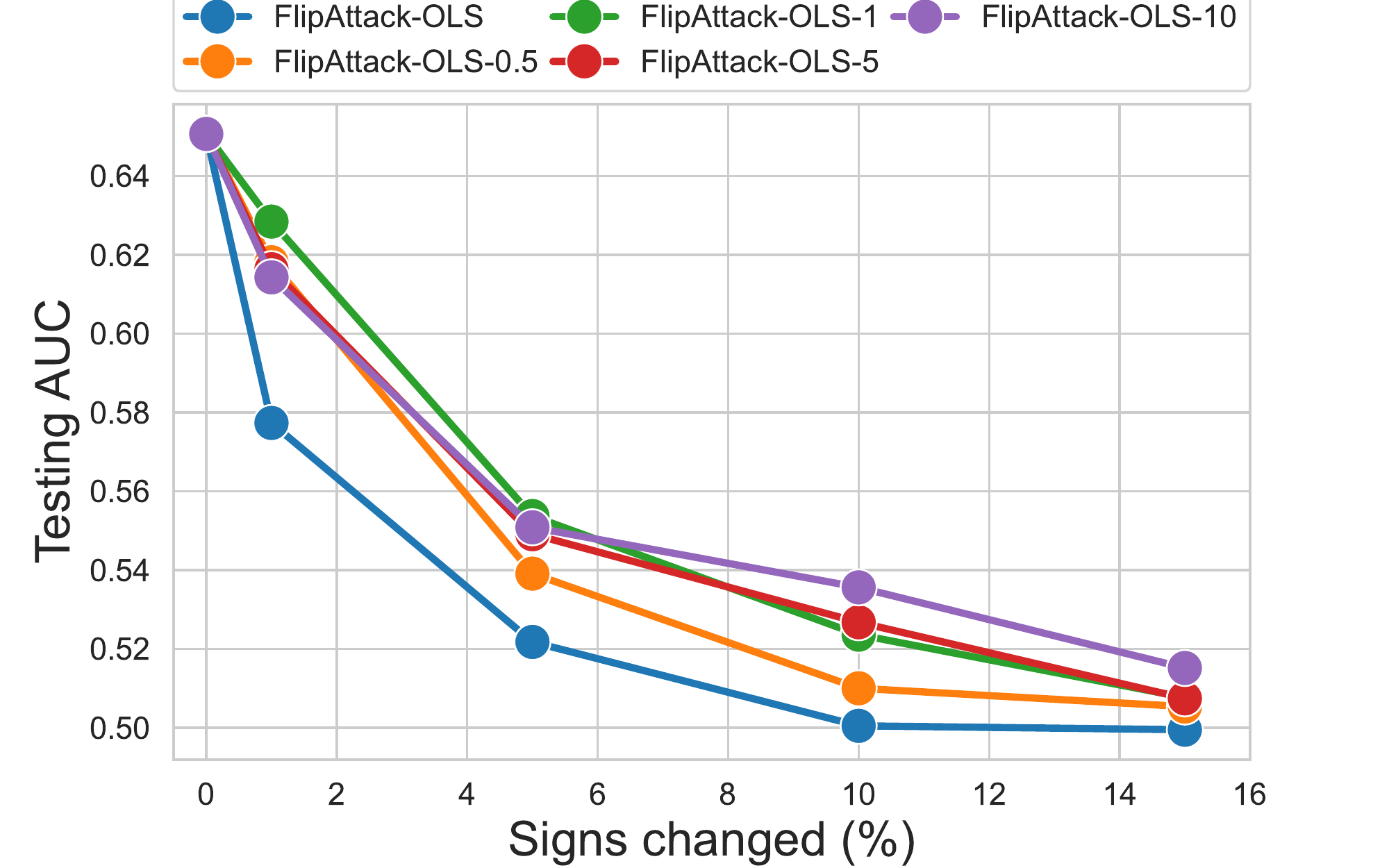}}
	\hfill
     \subfloat[\textsf{POLE}\label{1}]{%
       \includegraphics[width=0.5\linewidth]{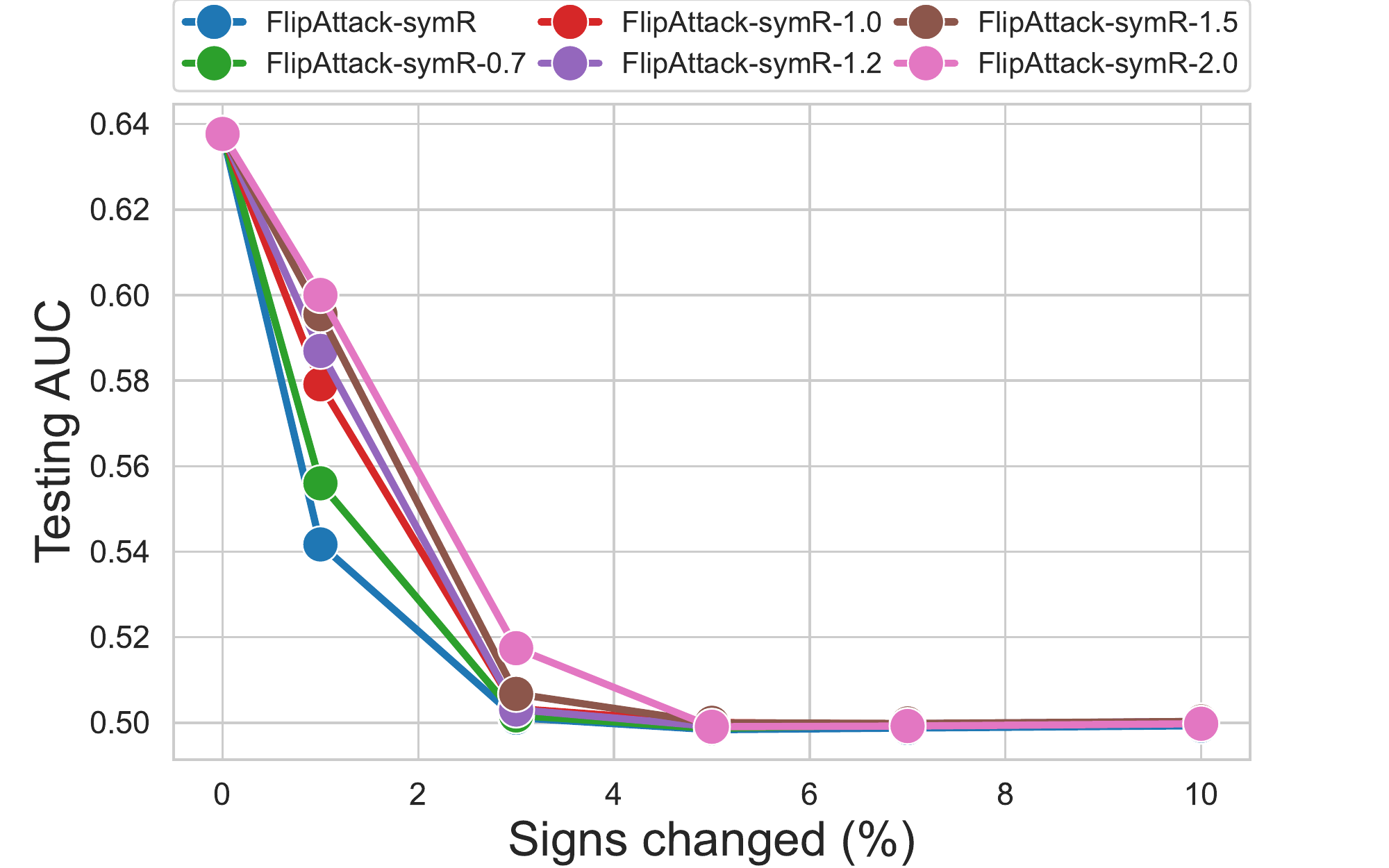}}
	\caption{The attack performances of refined attacks under different penalties.}
	\label{fig-tkg-auc}
\end{figure}

\subsubsection{Preserving attack performance}
We present the average AUC scores under refined attacks with various combinations of $\lambda$ and $\eta$ in Fig.~\ref{fig-tkg-auc}. The plausible result is that there exist combinations with which the refined attacks have almost the same attack performance as that of the basic attacks, even penalty terms are added. As expected, we can also observe the general trend that larger parameters (i.e., more penalty) will result in less effective attacks.

\subsubsection{Evading \textsf{Metric-View}}
In Tab.~\ref{tab-detect-OLS} and \ref{tab-detect-symR}, we show the trade-off between attack performance and secrecy with different configurations of the parameters $\lambda$ and $\eta$ for $\mathsf{FlipAttack}$-OLS and $\mathsf{FlipAttack}$-symR. Specifically, $\tau$ is the average AUC score under attack power $10\%$ for $\mathsf{FlipAttack}$-OLS and $5\%$ for $\mathsf{FlipAttack}$-symR, which is used as the mark for attack performance.

Our first observation is that tuning $\lambda$ along (set $\eta = 0$) can make $\mathsf{FlipAttack}$-OLS effectively evade \textsf{Metric-View} (the mean testing AUC drops from $0.982$ to $0.787$). This result coincides with that in Fig.~\ref{fig-balance-under-attack} where $\mathsf{FlipAttack}$-OLS will significant decreases $T(\mathcal{G})$ while having a relatively smaller impact on $Pol(\mathcal{G},t)$. Thus, imposing penalties on $T(\mathcal{G})$ alone is sufficient to evade the metric-based anomaly detector. 

Second, tunning $\eta$ along (set $\lambda = 0$) cannot effectively help $\mathsf{FlipAttack}$-symR to evade \textsf{Metric-View}. However, only tunning $\lambda$ can significantly degenerate the performance of the anomaly detection (mean AUC score drops from $0.94$ to $0.46$). These results show that even \textsf{Metric-View} is designed based on two features ($T(\mathcal{G})$ and $Pol(\mathcal{G},t)$), $T(\mathcal{G})$ (i.e., local structural balance) is the dominant one. In addition, we note that for all the configurations of $\lambda$ and $\eta$, there is little sacrifice on the attack performance (the worst case only increases from $0.5$ to $0.53$).

\begin{figure}[t]
	\centering
\subfloat[$\mathsf{FlipAttack}$-OLS\label{1a}]{%
       \includegraphics[width=0.5\linewidth]{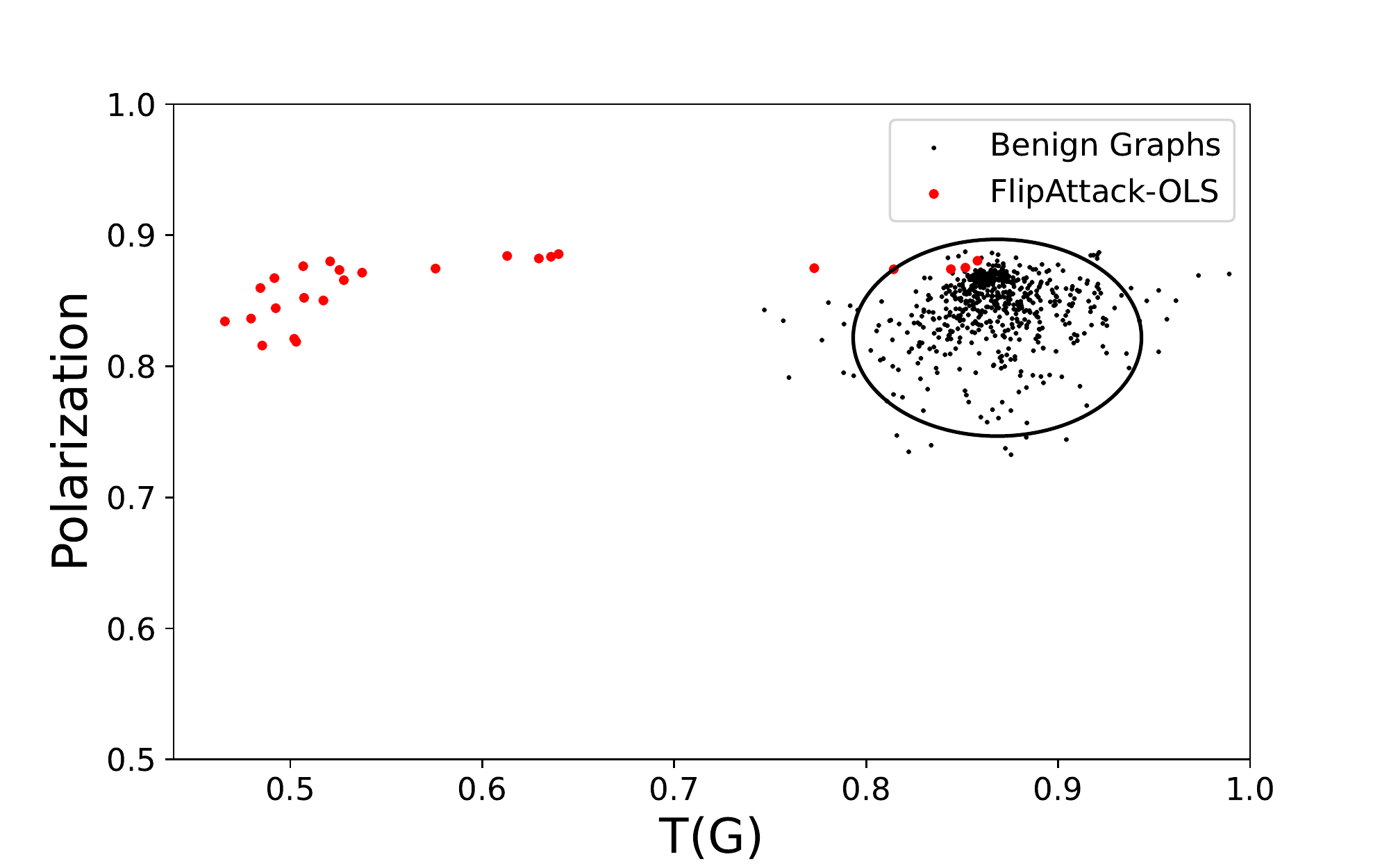}}
	\hfill
 \subfloat[$\mathsf{FlipAttack}$-OLS-$\lambda$\label{1a}]{%
       \includegraphics[width=0.5\linewidth]{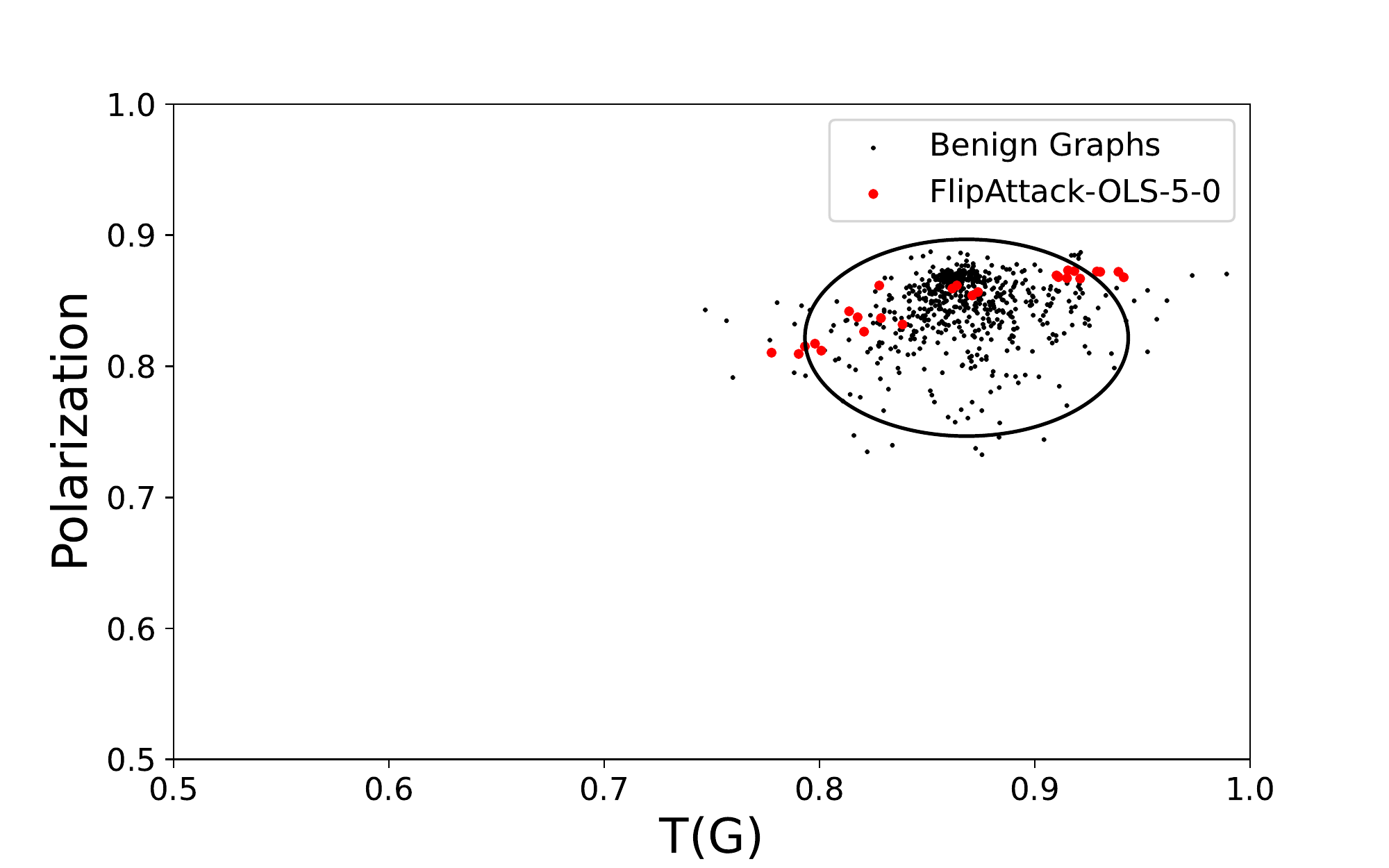}}
	\caption{(a) is the scatterplot of graph features on \textsf{Metric-View} for $\mathsf{FlipAttack}$-OLS; (b) is the scatterplot of graph features on \textsf{Metric-View} for $\mathsf{FlipAttack}$-OLS penalizing on $T(G)$.}
	\label{fig-scatter-ols}
\end{figure}

\begin{figure}[t]
	\centering
 \subfloat[$\mathsf{FlipAttack}$-symR\label{1a}]{%
       \includegraphics[width=0.5\linewidth]{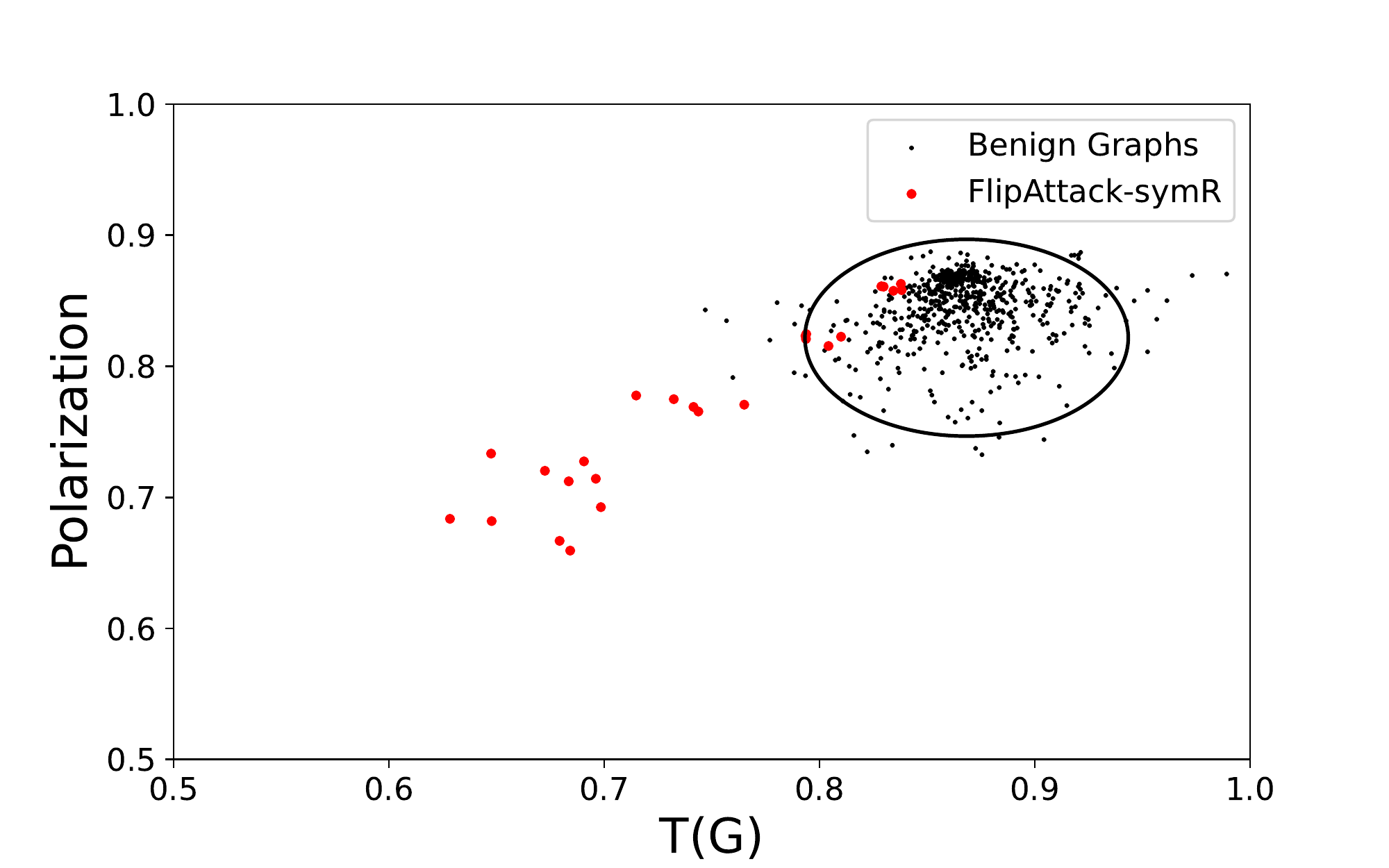}}
	\hfill
  \subfloat[$\mathsf{FlipAttack}$-symR-$\lambda$\label{1a}]{%
       \includegraphics[width=0.5\linewidth]{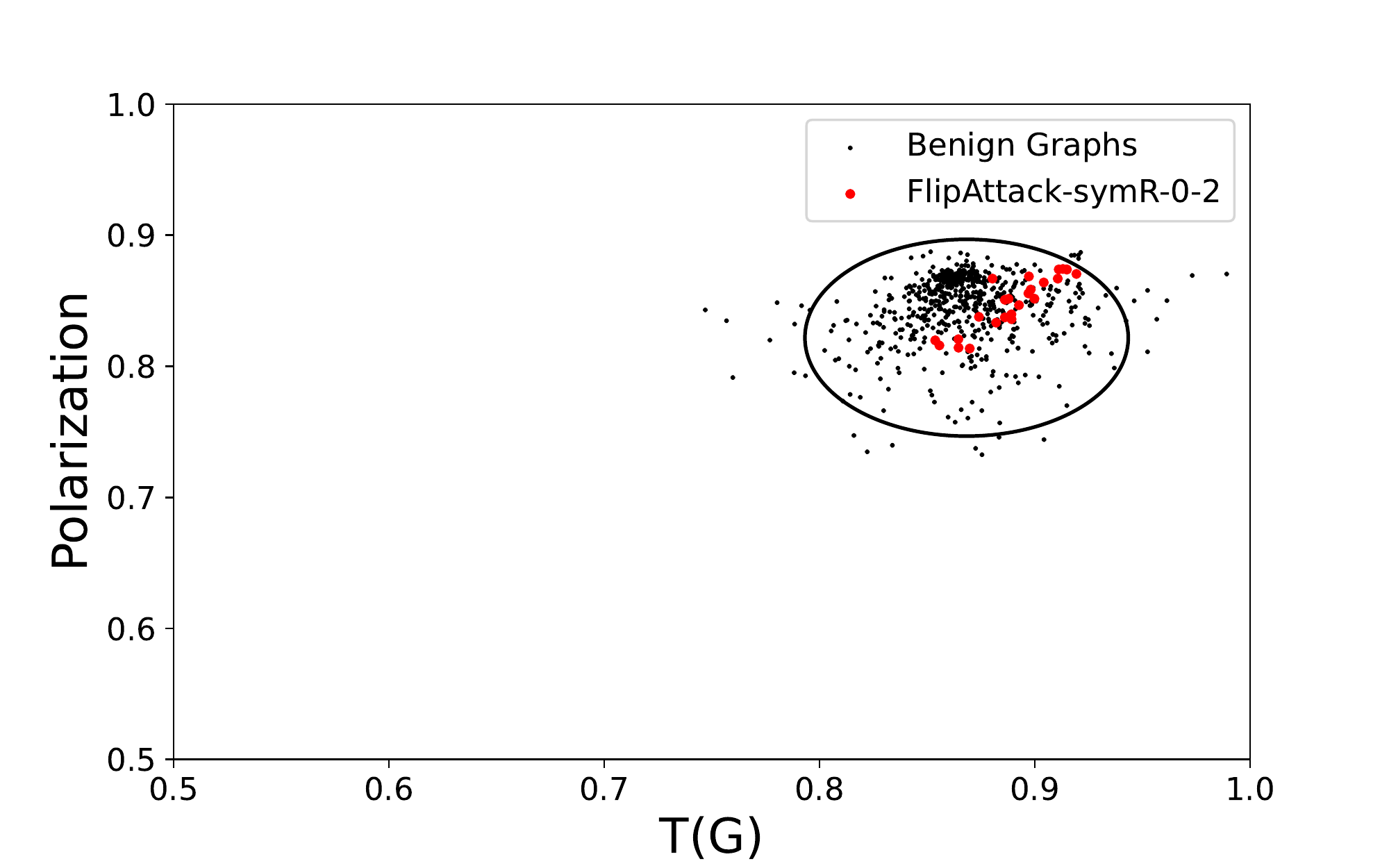}}
	\caption{(a) is the scatterplot of graph features on \textsf{Metric-View} for $\mathsf{FlipAttack}$-symR; (b) is the scatterplot of graph features on \textsf{Metric-View} for $\mathsf{FlipAttack}$-symR penalizing on $T(G)$.}
	\label{fig-scatter-symR}
\end{figure}

Figs.~\ref{fig-scatter-ols} and \ref{fig-scatter-symR} illustrate the basic attacks and secrecy-aware attacks for attacking \textsf{FeXtra} and \textsf{POLE}. As can be observed, only tuning $\lambda$ for $\mathsf{FlipAttack}$-OLS and $\mathsf{FlipAttack}$-symR pushes the abnormal data points into the decision boundary of the kernelized OCSVM, these results are in line with the quantitative analysis in Table~\ref{tab-detect-OLS} and Table \ref{tab-detect-symR}.

\subsubsection{Evading \textsf{TSVD-View}}
Unlike \textsf{Metric-View}, \textsf{TSVD-View} is both sensitive to $T(G)$ and $Pol(G,t)$ for attacking \textsf{FeXtra} (set $\lambda=2$ or $\eta=2$ will significantly decrease AUC scores $13.6\%$ and $33.4\%$). However, when attacking \textsf{POLE}, \textsf{TSVD-View} is more sensitive to penalizing $Pol(G,t)$. For example, if $\tau$ increases from $0$  to $2$, the mean AUC score decreases from $0.781$ to $0.675$.

\subsubsection{Evading \textsf{SGCN-View}}
We observe that both penalizing $T(\mathcal{G})$  and $Pol(G,t)$ for $\mathsf{FlipAttack}$-OLS and $\mathsf{FlipAttack}$-symR can effectively degrade the detection performance of \textsf{SGCN-View}. However, both $\mathsf{FlipAttack}$-OLS and $\mathsf{FlipAttack}$-symR are more sensitive to $\lambda$. Specially, tuning $\lambda=2$ for $\mathsf{FlipAttack}$-symR achieves the best evading performance (decrease mean AUC score from $0.856$ to $0.508$, a near fair toss). Intuitively, the \textsf{SGCN} more depends on the local structural balance to guide the node aggregate information along the balance and imbalance paths individually. Thus \textsf{SGCN-View} is more sensitive to the local structural balance.

\subsubsection{Evading $\mathsf{MvSGAD}$}
For ensemble learning, we observe that $\mathsf{MvSGAD}$ can achieve perfect detecting performance on basic attacks compared with the individual view. In consideration of the secrecy-aware attacks, $\mathsf{MvSGAD}$ is relatively sensitive to both $T(G)$ and $Pol(G,t)$. To be detailed, tuning $\lambda$ or $\eta$ from $0$ to $2$ can achieve $31.5\%$ decreasing percentage of mean AUC scores for attacking \textsf{FeXtra}, while the degeneration percentage of attacking \textsf{POLE} are $38.8\%$ and $43.5\%$. These phenomenons show that by penalizing on $T(G)$ and $Pol(G,t)$ indeed mitigate the side effects of the $\mathsf{FlipAttack}$ against \textsf{FeXtra} and \textsf{POLE}.

In summary, our comprehensive experiments demonstrate a few key insights. Firstly, imposing a penalty on the property chosen conflicting metrics could refine attacks such that they could mitigate the side effects of the poisoning attacks. Secondly, by properly choosing the configurations of penalty terms, the powerful attack detectors could be evaded with a high chance. Finally, there is indeed a trade-off between attack performance and evasion. Fortunately (or \textit{Unfortunately} for defenders), by choosing the proper parameters, we can achieve secrecy-aware attacks that can evade detection with a good chance while preserving attack performance.

\textbf{Discussion}\quad \textit{So, who will win this arms race between the attacker and the defender}? Can the attacker design stealthy attacks that can evade \textit{any} detectors? Or can the defender eventually find a magic detector that no attacks can evade? -- These are indeed some important open questions in this community. Our results provide some insights on attacks over signed graphs: in particular, how the poisoning attacks will change the graph data and what we should focus on when designing attack detectors. Such insights might contribute to this challenging topic of attack and defense over graphs.

\subsection{Attack Transferability}
\label{sec-transfer}
\begin{figure}[!t]
	\centering
     \subfloat[$\mathsf{FlipAttack}$-OLS-$\lambda$\label{OLS-lamb}]{%
       \includegraphics[width=0.5\linewidth]{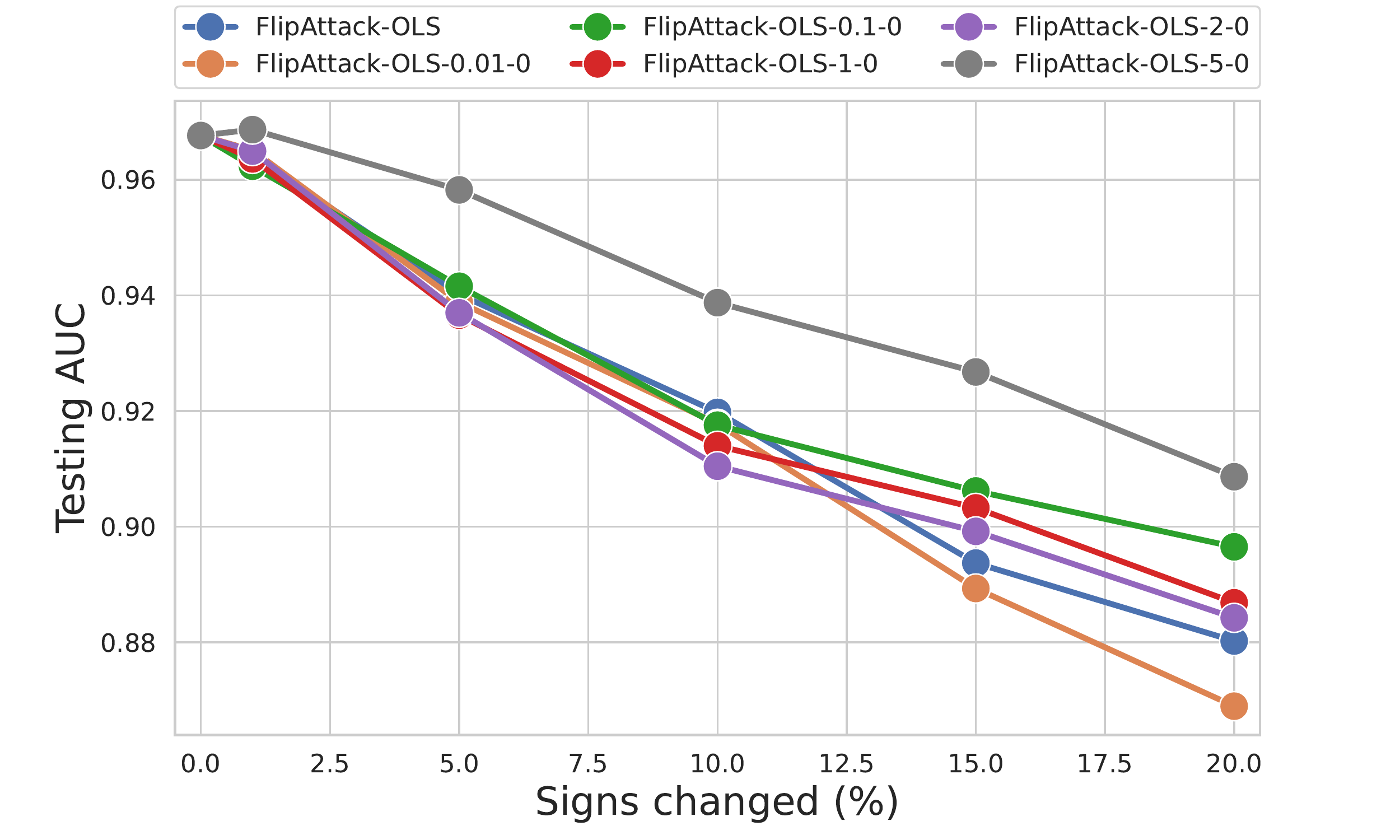}}
	\hfill
      \subfloat[$\mathsf{FlipAttack}$-OLS-$\eta$\label{OLS-eta}]{%
       \includegraphics[width=0.5\linewidth]{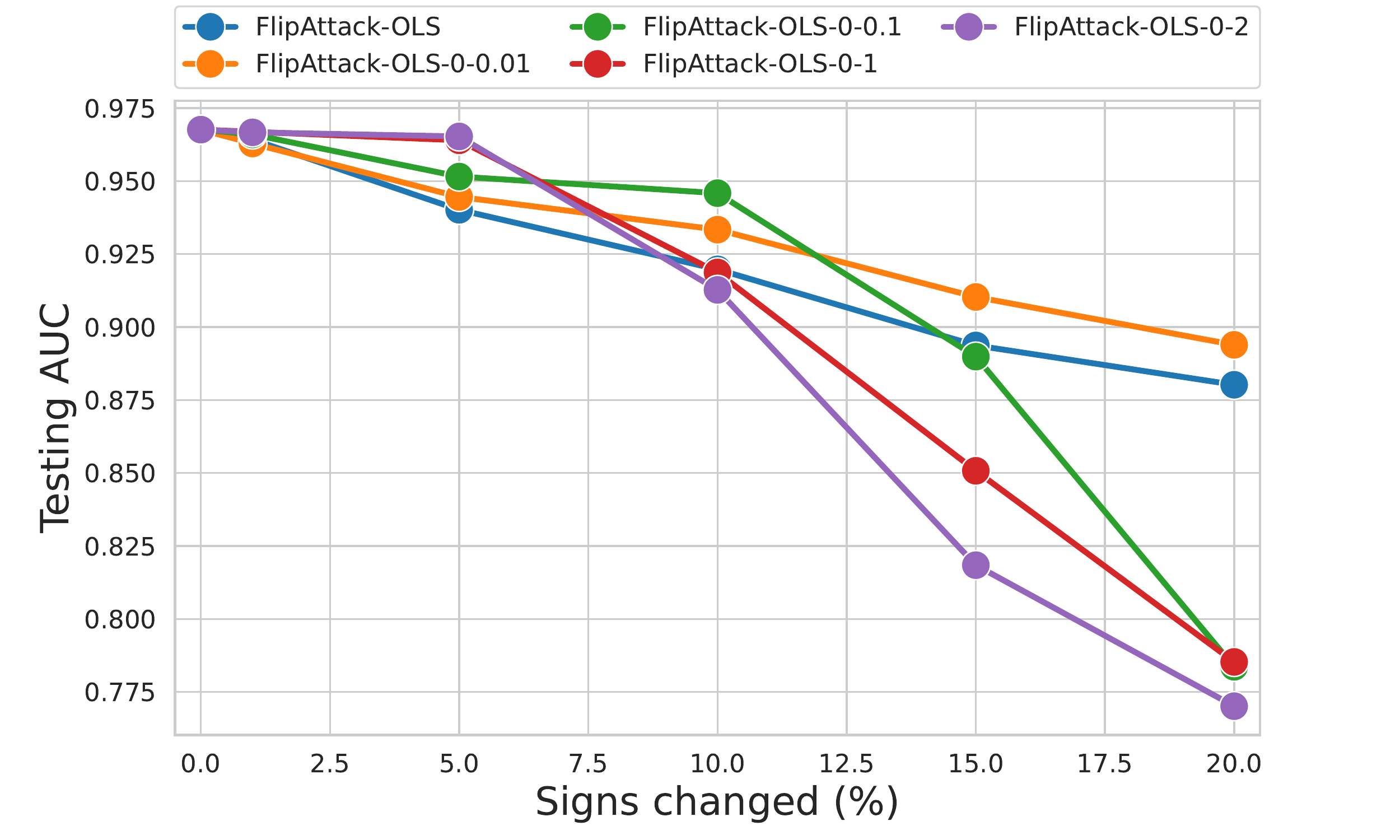}}
	\hfill
\subfloat[$\mathsf{FlipAttack}$-symR-$\lambda$\label{symR-lamb}]{%
       \includegraphics[width=0.5\linewidth]{figure/transfer_SGCN_OLS_Polarization.pdf}}
	\hfill
 \subfloat[$\mathsf{FlipAttack}$-symR-$\eta$\label{symR-eta}]{%
       \includegraphics[width=0.5\linewidth]{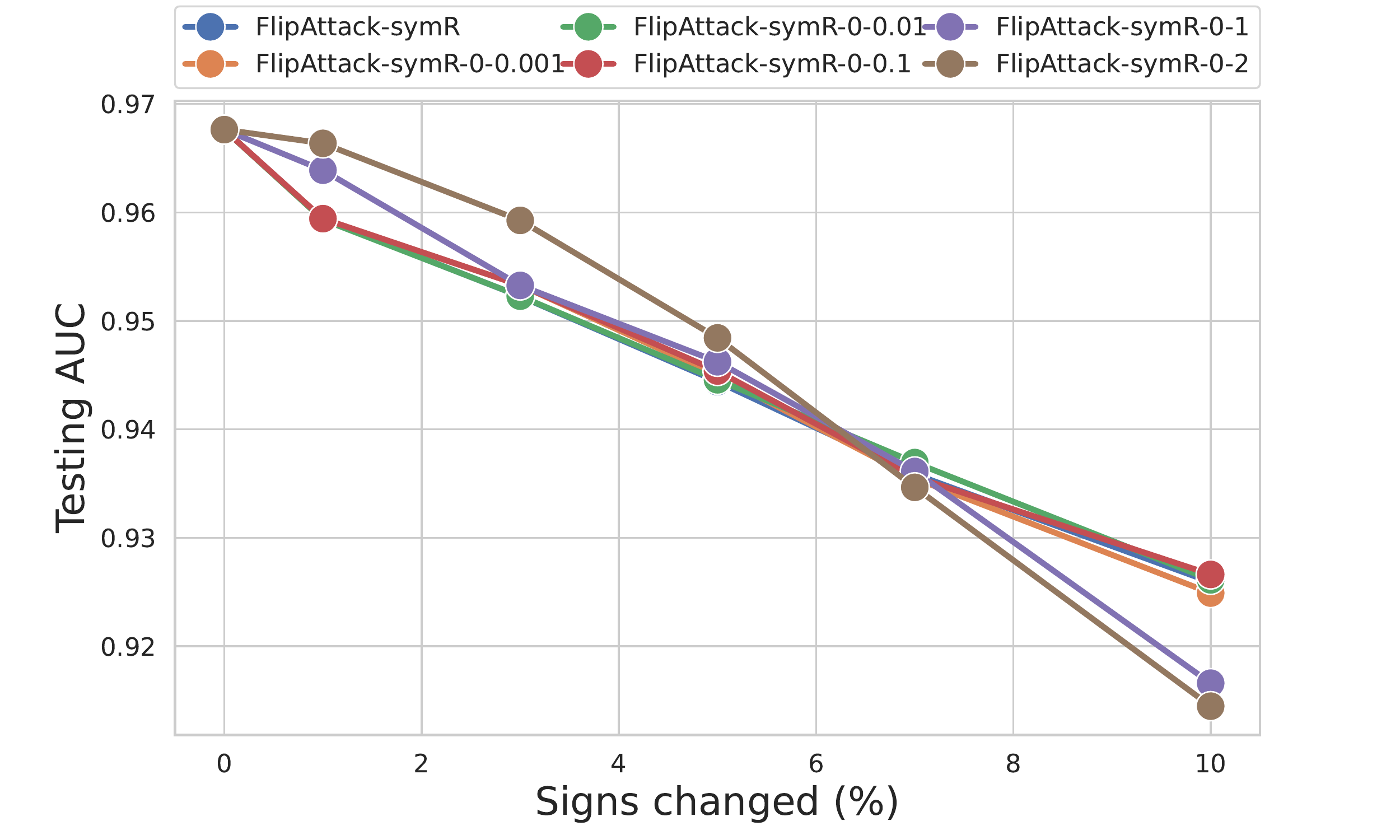}}
	\caption{Transfer attack to \textsf{SGCN} in poisoning manner. $\mathsf{FlipAttack}$-OLS-$\lambda$ means $\mathsf{FlipAttack}$-OLS penalizing $T(G)$ while $\mathsf{FlipAttack}$-OLS-$\eta$ is penalizing on $Pol(G,t)$.}
	\label{fig-auc-transfer-sgcn}
\end{figure}

\begin{figure}[!t]
	\centering
 \subfloat[$\mathsf{FlipAttack}$-OLS-$\lambda$\label{OLS-lamb}]{%
       \includegraphics[width=0.5\linewidth]{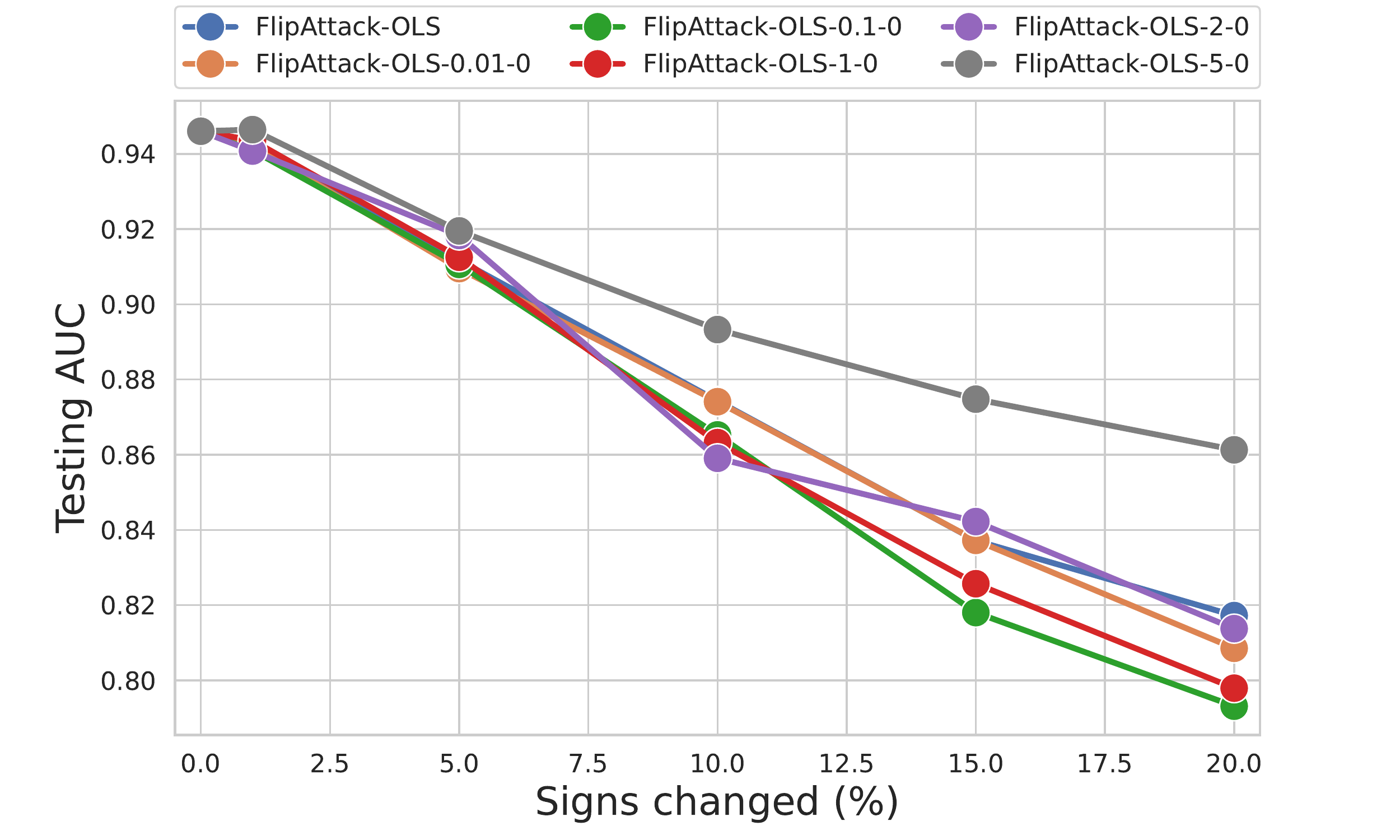}}
	\hfill
 \subfloat[$\mathsf{FlipAttack}$-OLS-$\eta$\label{OLS-eta}]{%
       \includegraphics[width=0.5\linewidth]{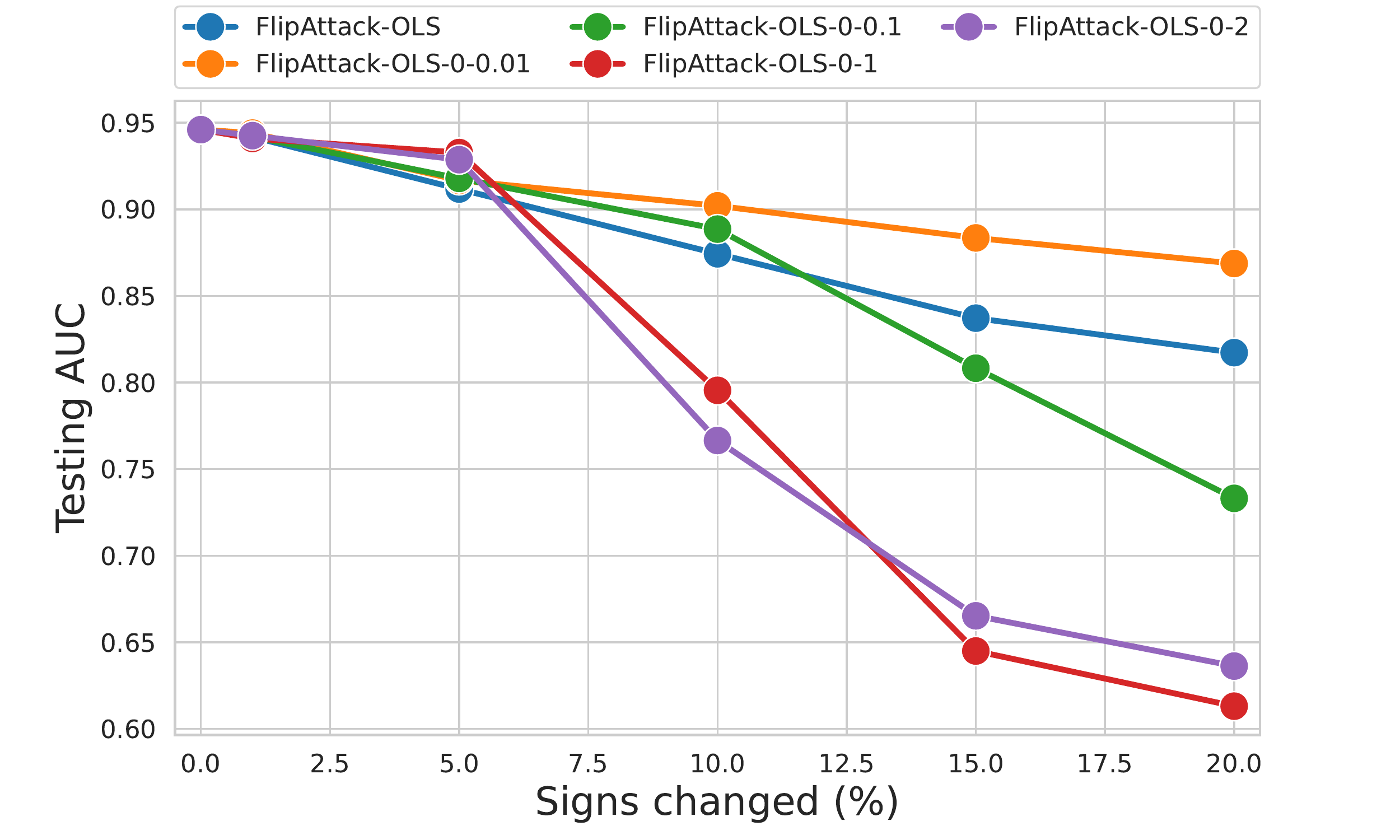}}
	\hfill
  \subfloat[$\mathsf{FlipAttack}$-symR-$\lambda$\label{symR-lamb}]{%
       \includegraphics[width=0.5\linewidth]{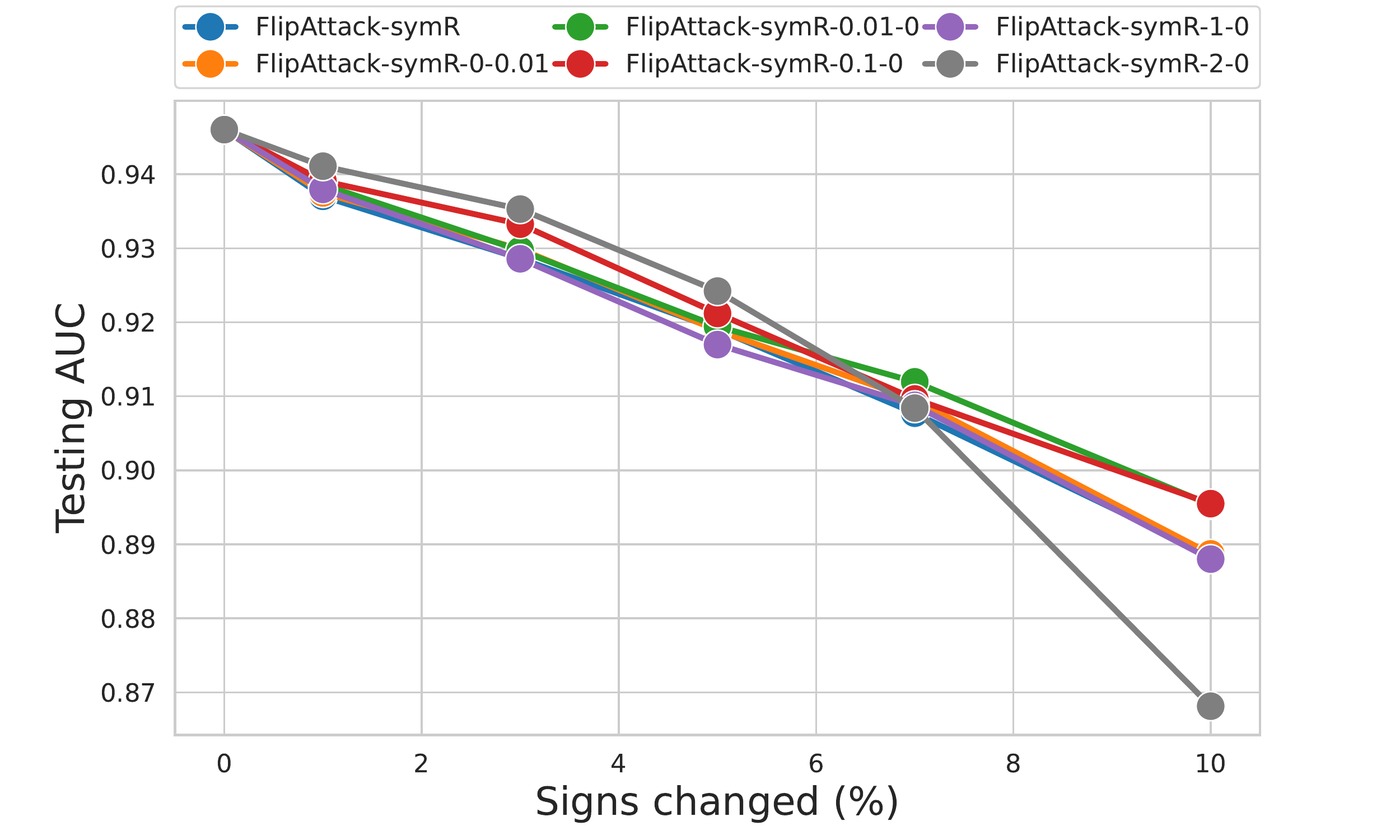}}
	\hfill
   \subfloat[$\mathsf{FlipAttack}$-symR-$\eta$\label{symR-eta}]{%
       \includegraphics[width=0.5\linewidth]{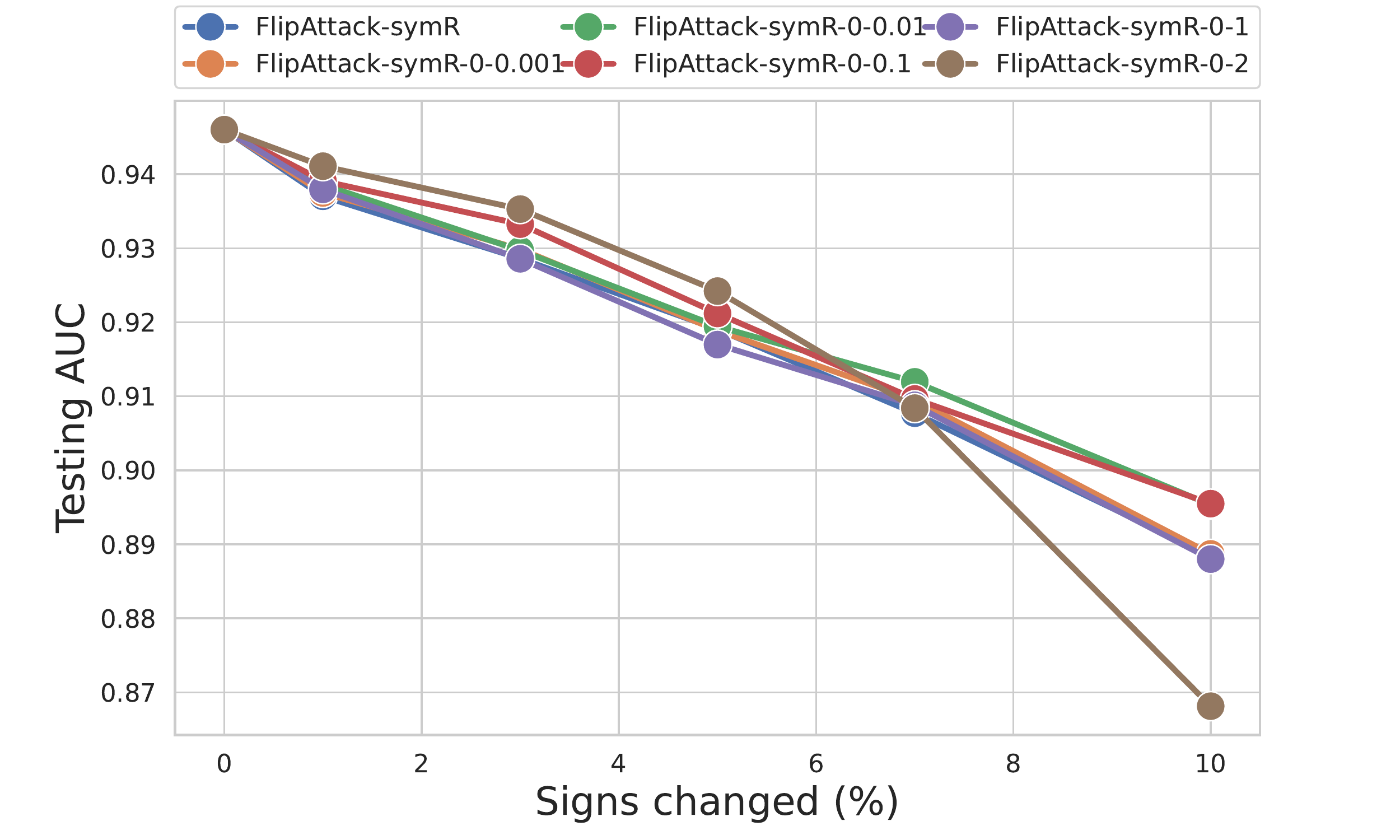}}
	\caption{Transfer attack to \textsf{SNEA} in poisoning manner. $\mathsf{FlipAttack}$-symR-$\lambda$ means $\mathsf{FlipAttack}$-symR penalizing $T(G)$ while $\mathsf{FlipAttack}$-symR-$\eta$ is penalizing on $Pol(G,t)$.}
	\label{fig-auc-transfer-snea}
\end{figure}

Recently, there exists a surge of using GNN-based models for link sign prediction whose aggregation mechanism is especially designed for signed graphs. In practice, the choice of the prediction models could remain unknown to the attacker. Thus, there is a need to test the transferability of attacks, i.e., the ability of an attack to mislead the trust prediction of a model that it is not designed for. To this end, we evaluate the transferability of our proposed attacks against two representative GNN-based models: \textsf{SGCN} \cite{sgcn} and \textsf{SNEA} \cite{snea}. Specifically, \textsf{SGCN} utilizes the local balance theory and allows the center node to aggregate neighbor's information along the balance path and imbalance path separately. \textsf{SNEA} takes a further step to incorporate the signed convolutional layer with the graph attention mechanism to boost the prediction performance. In our experiment, we investigate whether the poisoned graphs obtained from  $\mathsf{FlipAttack}$ can also degenerate the link sign prediction performance of \textsf{SGCN} and \textsf{SNEA}. In detail, we feed the poisoned graphs with different attacking powers into \textsf{SGCN} and \textsf{SNEA} and retrain the target model (also in a poisoning manner). 
Then, we evaluate the link sign prediction performance of these two models using the testing AUC scores. The experiment results are shown in Fig.~\ref{fig-auc-transfer-sgcn} and \ref{fig-auc-transfer-snea}. The results show that attacking \textsf{FeXtra} and \textsf{POLE} can both degrade the performance of the GCN-based models especially \textsf{SNEA}. Intuitively, attacking \textsf{FeXtra} and \textsf{POLE} can effectively destroy the balance property of the signed graphs, leading to inaccurate node aggregation path in the signed convolutional layer and wrong prediction. Specially, $\mathsf{FlipAttack}$-OLS with $\eta=1$ gains the best attacking performance under the attacking power equal to $20\%$, leading to $36.8\%$ decreasing percentage of the AUC score on testing data.

\section{Conclusion}
\label{sec-conclusion}
In this paper, we study adversarial attacks against trust prediction in signed graphs. We propose four basic attacks that can effectively downgrade the classification performances for two typical machine learning models \textsf{FeXtra} and \textsf{POLE}. However, we show that these basic attacks would inevitably break the structural semantics of signed graphs, making them prone to be detected. We thus further devise a joint-optimization approach to realize refined attacks that have this nice property: they can evade attack detectors with high probability while sacrificing little attack performance. In other words, the refined attacks are secrecy-aware. Our results mark a critical step towards more practical attacks. 

\bibliographystyle{IEEEtranN}

\bibliography{citation}
\end{document}